%% file: notes_ArXiv_v2.tex
\documentclass[12pt, a4paper, twoside]{article}

\usepackage[a4paper, left=3cm,right=2cm]{geometry}
\usepackage{hyperref}
\usepackage{amsfonts}
\usepackage{amsmath}
\usepackage{empheq}
\usepackage{setspace}
\usepackage{color}
\usepackage{esint}
\usepackage{slashed}
\usepackage{enumitem}
\usepackage[british]{babel}
\usepackage{hyphenat}

\numberwithin{equation}{section}

\usepackage{cite}
\usepackage{graphicx, import}
\usepackage{calc}
\usepackage[font=small, font+=it, format=hang, margin={1cm, 1cm}]{caption}
\graphicspath{{figures/}}

\newcommand{\dsp}{\displaystyle}

\newcommand{\hs}[1]{\hspace{#1 em}}
\newcommand{\bflr}{\begin{flushright}}
\newcommand{\eflr}{\end{flushright}}
\newcommand{\bc}{\begin{center}}
\newcommand{\ec}{\end{center}}
\newcommand{\ben}{\begin{enumerate}}
\newcommand{\een}{\end{enumerate}}

\newcommand{\be}{\begin{equation}}
\newcommand{\ee}{\end{equation}}
\newcommand{\ba}{\begin{array}}
\newcommand{\ea}{\end{array}}
\newcommand{\ct}{\cite}

\newcommand{\ag}{\alpha}
\newcommand{\bg}{\beta}
\newcommand{\gam}{\gamma}
\newcommand{\del}{\delta}

\newcommand{\ve}{\varepsilon}

\newcommand{\thg}{\theta}
\newcommand{\kg}{\kappa}
\newcommand{\lb}{\lambda}
\newcommand{\sg}{\sigma}
\newcommand{\rg}{\rho}

\newcommand{\og}{\omega}
\newcommand{\Gam}{\Gamma}
\newcommand{\Del}{\Delta}
\newcommand{\Fg}{\Phi}
\newcommand{\Sg}{\Sigma}
\newcommand{\Og}{\Omega}
\newcommand{\Lb}{\Lambda}

\newcommand{\bfk}{{\bf k}}

\newcommand{\lh}{\left(}
\newcommand{\rh}{\right)}
\newcommand{\ld}{\left.}
\newcommand{\rd}{\right.}

\newcommand{\der}{\partial}

\newcommand{\de}{\mbox{d}}
\newcommand{\e}{\mbox{e}}
\newcommand{\phant}{\phantom{a}}

\title{Lecture notes on dynamical spacetime and gravitational waves}
\begin{document}

\begin{titlepage}

\begin{flushright}\vspace{-3cm}KCL-PH-TH/2017-06\end{flushright}\vspace{0.5cm}

\begin{center}{\large{\bf Field theoretical approach to gravitational waves}}
\end{center}
\vspace{10mm}
\centerline{\normalsize{\bf{M.~de Cesare\footnote{marco.de\_cesare@kcl.ac.uk}}, \bf{R.~Oliveri\footnote{roliveri@ulb.ac.be}}, 
\bf{J.W.~van Holten\footnote{v.holten@nikhef.nl}}}}
\vspace{2em}
\begin{center}
\textit{$\phant^{1}$Theoretical Particle  Physics  and  Cosmology Group,
Department  of  Physics,  King's  College  London,
University of London,  Strand,  London, WC2R 2LS, U.K.\\~\\$\phant^{2}$Service de Physique Th{\'e}orique et Math{\'e}matique,\\ Universit{\'e} Libre de Bruxelles and International Solvay Institutes,\\ Campus de la Plaine, CP 231, B-1050 Brussels, Belgium\\~\\
$\phant^{3}$Nikhef, Science Park 105, 1098 XG Amsterdam, Netherlands}
\end{center}

\vspace{5mm}
\normalsize
\bigskip\medskip
\textit{
}

\vspace{5mm}

\begin{abstract}
\noindent 
{The aim of these notes is to give an accessible and self-contained introduction to the theory of gravitational waves 
as the theory of a relativistic symmetric tensor field in a Minkowski background spacetime.  This is the approach of a particle 
physicist: the graviton is identified with a particular irreducible representation of the Poincar{\'e} group, corresponding to vanishing mass and spin two. It is shown how to construct an action functional giving the linear dynamics of gravitons, and how General Relativity can be obtained from it. The Hamiltonian formulation of the linear theory is examined in detail. We study the emission of gravitational waves and apply the results to the simplest case of a binary Newtonian system.
}
\end{abstract}

\begin{center}
\end{center}

\end{titlepage}

\tableofcontents

\section*{Introduction}
Gravitational waves have recently become a hot topic in physics since their direct observation by the LIGO experiment \cite{Abbott:2016blz,Abbott:2016nmj}. Two signals have been detected, labelled GW150914 and GW151226, corresponding to the inspiralling and subsequent merging of binary black holes systems. The discovery opens a new window for the observations of extreme astrophysical phenomena and marks the beginning of the era of gravitational wave astronomy. Moreover, future observations can potentially have important consequences also for cosmology. The discovery comes nearly one century after gravitational waves were first predicted by Einstein \cite{Einstein:1916cc}, in 1916, one year after he wrote down the gravitational field equations \cite{Einstein:1916vd}. Gravitational waves are one of the most remarkable predictions of Einstein's theory of General Relativity, and the one for 
which it proved hardest to find direct experimental confirmation. The reason for this is the very weak coupling of systems of ordinary masses and length scales to gravity, as expressed by the smallness of the gravitational constant. The emission of a large amount of energy in the form of gravitational radiation requires very compact systems, such as neutron stars and black holes or the occurrence of extreme astrophysical phenomena, such as supernovae.

As a starting point we take the point of view of Special Relativity, where spacetime is flat and described by the Minkowski geometry; it merely represents a local stage for physical processes to take place on. Flat spacetime is not only the simplest background one can consider, but it is also quite special due to its high degree of symmetry. In fact, it has a group of global isometries: the Poincar{\'e} group. Field theories compatible with Special Relativity must respect such invariance of the background. Furthermore, the very definition of a particle in relativistic quantum field theory relies on the Poincar{\'e} group; they are identified with its irreducible representations and labelled by two numbers, namely the spin and the mass. We introduce gravitational waves as some particular irreducible representations of the Poincar{\'e} group, massless with spin two, and build an action principle describing their dynamics. This alternative approach is complementary and equivalent to the standard one, where gravitational waves are found as solutions of the linearised Einstein's equations. It is similar to that of Feynman's lectures on gravity, held at Caltech in 1962-63 \cite{Feynman:1996kb}, and Veltman's lectures held at Les Houches \cite{Veltman:1976}. The fact that the gravitational interaction can be described as mediated by a spin-two field is understood on the basis of the universality of gravity, which means that it must couple to all forms of energy.
Its masslessness is due to the long range of the interaction.

These notes extend the contents presented in \cite{vanHolten:2016avu} and deal with the topic in more detail.
They are organised as follows.
The study of irreducible representations of the Poincar{\'e} group is covered in the Sec.~\ref{Sec 1}, where we show in particular how they can be extracted from tensor representations, which are in general reducible. Tensors can therefore be decomposed in terms of irreducible representations corresponding to different spins. A correct kinematical description of the gravitational perturbations requires identifying such spurious, lower spin components, and projecting them out. While this is straightforward at the kinematical level, following the procedure laid out in Sec.~\ref{Sec 1}, it leads to subtle issues when formulating the dynamics, as discussed in Sec.~\ref{Sec 2}. The analysis must be carried out on a case-by-case basis for each tensor representation. To make things simple and to be as general as possible, we consider the problem of formulating the dynamics of a massive rank-two tensor field. We require that all but the spin-two component are non-dynamical, hence not propagating on spacetime. The dynamics of the gravitational perturbations can then be recovered by formally setting the mass parameter to zero. In Sec.~\ref{Sec 3}, the action is recast in the form of the Fierz-Pauli action for linear massive gravity, which agrees with linearised General Relativity for vanishing mass. The theory has a local symmetry parametrised by a vector field, which displays remarkable similarities with the gauge symmetries familiar from Yang-Mills theories. In fact, this can be recognised as the linear version of the diffeomorphism invariance of the Einstein-Hilbert action. The full nonlinear theory can then be recovered by 
the Noether construction, extending the action and gauge transformations order-by-order in the gravitational coupling parameter. In Sec.~\ref{Sec 4}, we introduce the Hamiltonian formalism to study the dynamics of gravitational waves. This formalism allows one to recognise that the dynamics of the gravitational perturbations is constrained. Specifically, the number of dynamical degrees of freedom of the gravitational field is more transparent in this formalism. The occurrence of constraints in the Hamiltonian formulation of the dynamics is a consequence of gauge invariance of the theory and it is common to all fundamental interactions. It amounts to the fact that one is introducing redundancies into the physical description of the system. In the full nonlinear theory of gravity, such redundancies can be identified with the existence of different systems of local coordinates which must all be physically equivalent. Physical configurations are therefore identified with classes of gauge equivalent solutions of the dynamics. A more practical way to describe them in the linearised theory is to pick a representative from each class, by means of a gauge fixing procedure which is familiar from electromagnetism. We explain in detail how the gauge freedom can be fixed, considering in particular the TT gauge relevant for applications to gravitational waves. In Sec.~\ref{Sec 5}, we write down a continuity equation for the energy-momentum flux carried by gravitational waves and discuss their emission by sources. The quadrupole formula is derived. Finally, in Sec. \ref{Sec 6}, we consider the particularly relevant case of the emission of gravitational radiation from a system of Newtonian binaries, and discuss other sources of gravitational waves. In the Outlook, Sec.~7, we briefly summarise our work and put it into the more general context of research in gravitational waves.\\

\newpage

\section*{Conventions}
We consider units in which $\hbar=c=1$. The Minkowski metric is
\be \nonumber
\eta_{\mu\nu}=\begin{pmatrix} -1 &\phant&\phant&\phant\\ \phant& 1& \phant&\phant\\ \phant&\phant&1&\phant\\ \phant&\phant&\phant&1\end{pmatrix}.
\ee
Spacetime tensor components are denoted by Greek letters, whereas letters from the Latin alphabet are used for spatial components.
\newpage

\section{Representations of the Poincar\'e group} \label{Sec 1}
In this section, we study the representation theory of the group of global isometries of Minkowski spacetime, the Poincar\'e group. Considering some particularly relevant examples, we show how irreducible representations can be extracted from tensor representations and how they are classified according to the mass and spin of the fields.
 
According to Special Relativity, the laws of Physics must be the same for any inertial observer. In other words, they must be invariant in form under special Lorentz transformations (also called boosts), rotations, translations and any arbitrary compositions of the above. These transformations together generate what is called the Poincar\'{e} group, which therefore represents the fundamental symmetry of any relativistic theory. The geometric setting in Special Relativity is fixed at the outset and is given by Minkowski spacetime, whose metric tensor will be denoted as $\eta_{\mu\nu}$. Inertial observers are associated to orthonormal frames. The Poincar\'e group maps the class of inertial frames into itself. This statement is equivalent to saying that the metric $\eta_{\mu\nu}$ is invariant under the action of the group, whose elements are therefore isometries of flat spacetime.

Since the spacetime is fixed in Special Relativity, all the interesting physics lies in the dynamics of particles on the inert flat background. At a fundamental level, all particles and fundamental interactions are described in terms of fields defined on Minkowski spacetime. In a (special) relativistic theory, each field must respect the symmetries of the background, and therefore belongs to a certain (irreducible) representation of the Poincar\'{e} group. It is therefore of primary importance to classify these representations and understand their content in terms of kinematical degrees of freedom before turning to the construction of their dynamics. The gravitational field will be understood in the weak field limit as one particular such representations.

For our purposes it is more convenient to work with infinitesimal rather than finite symmetry transformations. This is the same as considering the Lie algebra of the Poincar\'{e} group instead of the group itself. The generators of the Lie algebra obey the following commutation relations 
\be
\ba{l}
\left[ P_{\mu}, P_{\nu} \right] = 0, \hs{1} \left[ J_{\mu\nu}, P_{\lb} \right] = \eta_{\nu\lb} P_{\mu} - \eta_{\mu\lb} P_{\nu}, \\
 \\
\left[ J_{\mu\nu}, J_{\kg\lb} \right] = \eta_{\nu\kg} J_{\mu\lb} - \eta_{\nu\lb} J_{\mu\kg} - \eta_{\mu\kg} J_{\nu\lb} 
 + \eta_{\mu\lb} J_{\nu\kg}. 
\ea
\label{1.1}
\ee
Here $P_{\mu}$ is the the generator of spacetime translations, whereas $J_{\mu\nu}$ generates Lorentz transformations (boosts and rotations). Observe that we take the generators to be anti-hermitean.

Considering a scalar field $F(x)$, the general expression of an infinitesimal Poincar\'{e} transformation  with 
real parameters $\ag^{\mu}$ and $\og^{\mu\nu} = - \og^{\nu\mu}$ reads as follows
\be
\del F = \lh \ag^{\mu} P_{\mu} + \frac{1}{2}\, \og^{\mu\nu} J_{\mu\nu} \rh F.
\label{1.2}
\ee
Translations, defining the nilpotent part of the algebra, always act only on the argument $x^{\mu}$ of the field, 
whereas Lorentz transformations and rotations act both on the spacetime argument and on field components. 
For this reason it is convenient to split the latter into an orbital part $M_{\mu\nu}$ and a spin part $\Sg_{\mu\nu}$ acting on the internal space only, in analogy with what one does with the generators of rotations in ordinary Quantum Mechanics:
\be
J_{\mu\nu} = M_{\mu\nu} + \Sg_{\mu\nu}.
\label{1.3}
\ee
The orbital and spin operators both satisfy the commutation relations of the Lorentz algebra separately
\be
\ba{rcl}
\left[ M_{\mu\nu}, M_{\kg\lb} \right] & = & \eta_{\nu\kg} M_{\mu\lb} - \eta_{\nu\lb} M_{\mu\kg} - \eta_{\mu\kg} M_{\nu\lb} 
 + \eta_{\mu\lb} M_{\nu\kg}, \\
  & & \\
\left[ \Sg_{\mu\nu}, \Sg_{\kg\lb} \right] & = & \eta_{\nu\kg} \Sg_{\mu\lb} - \eta_{\nu\lb} \Sg_{\mu\kg} - \eta_{\mu\kg} \Sg_{\nu\lb} 
 + \eta_{\mu\lb} \Sg_{\nu\kg}.
\ea
\label{1.4}
\ee

The generators of translations $P_{\mu}$ and the orbital Lorentz transformations $M_{\mu\nu}$ have a representation in terms of differential operators acting on fields defined on spacetime:
\be
P_{\mu} = \der_{\mu}, \hs{2} M_{\mu\nu} = x_{\mu} \der_{\nu} - x_{\nu}\der_{\mu}.
\label{1.5}
\ee
It is easy to check that they satisfy the algebra given by Eqs.~(\ref{1.1}) and (\ref{1.4}). For scalar fields, which 
have no spacetime components and carry no spin, this is enough to completely specify a representation of the 
Poincar\'{e} algebra. However, for fields with spacetime components, such as a vector field $A_{\mu}$ or a 
tensor field $A_{\mu\nu}$, the action of the spin operators $\Sigma_{\mu\nu}$ is non-trivial.
 
For a consistent implementation of the Poincar\'e algebra, Eq.~(\ref{1.1}), the operators $\Sg_{\mu\nu}$ have to commute with the translation operators and the orbital Lorentz transformations
\be
\left[ \Sg_{\mu\nu}, P_{\lb} \right] = 0, \hs{2} \left[ \Sg_{\mu\nu}, M_{\kg\lb} \right] = 0,
\label{1.6}
\ee 
thus defining a separate finite-dimensional representation of the Lorentz algebra as in Eq.~(\ref{1.4}). The generators 
$\Sg_{\mu\nu}$ act in the \emph{internal} space of the physical system considered, which describes the spin degrees of freedom and is labelled by discrete indices (spacetime or spinorial\footnote{We will not be concerned with the latter here.}).  For instance, the representation of the spin generators on vector fields is
\be 
\lh \Sg_{\mu\nu} \rh_{\ag}^{\phant \bg} = \eta_{\mu\ag} \del_{\nu}^{\phant\bg} - \eta_{\nu\ag} \del_{\mu}^{\phant\bg}, \hs{1}
\del A_{\ag} = \frac{1}{2}\, \og^{\mu\nu} \lh \Sg_{\mu\nu} \rh_{\ag}^{\phant\bg} A_{\bg} = \og_{\ag}^{\phant\bg} A_{\bg}.
\label{1.7}
\ee
When considering representations of spin generators on rank-two tensors $A_{\ag\bg}$, these are simply 
obtained by a straightforward linear extension of the above transformations
\be\label{eq:LorentzActingOnTwoTensors}
\lh \Sg_{\mu\nu} \rh_{\ag\bg}^{\phant\phant\gam\del} = 
    \lh \eta_{\mu\ag} \del_{\nu}^{\phant\gam} - \eta_{\nu\ag} \del_{\mu}^{\phant\gam} \rh \del_{\bg}^{\phant\del} 
    + \del_{\ag}^{\phant\gam} \lh \eta_{\mu\bg} \del_{\nu}^{\phant\del} - \eta_{\nu\bg} \del_{\mu}^{\phant\del} \rh.
\ee 
It is easy to check that both expressions in Eq.~\eqref{1.7} and Eq.~\eqref{eq:LorentzActingOnTwoTensors} obey the commutation relation Eq.~\eqref{1.4}.
Eq.~\eqref{eq:LorentzActingOnTwoTensors} might be modified with appropriate symmetrisation or anti-symmetrisation of the pairs of indices $(\ag\bg)$ and $(\gam \del)$, when necessary. In fact, as we will see later, some physical fields are described by tensor fields with special symmetry properties, which must be preserved by the action of the Poincar\'{e} group. In particular, the graviton, \emph{i.e.} the field representing gravitational waves, is represented by a symmetric rank-two tensor with zero trace. 
It follows from Eq.~(\ref{eq:LorentzActingOnTwoTensors}) that the action of the spin part of the Lorentz transformations on a rank-two tensor field can be written as follows
\be
\del A_{\ag\bg} = \frac{1}{2}\, \og^{\mu\nu} \lh \Sg_{\mu\nu} \rh_{\ag\bg}^{\phant\phant\gam\del} A_{\gam \del} 
 = \og_{\ag}^{\phant\gam} A_{\gam\bg} + \og_{\bg}^{\phant\gam} A_{\ag\gam}.
\label{1.9}
\ee

It is well known that the rotation group in three spatial dimensions is a subgroup of the Lorentz group, generated by the Lie algebra elements $J_{ij}$ $(i,j = 1,2,3)$, which satisfy the following commutation relations: 
\be
\left[ J_{ij} , J_{kl} \right] = \del_{jk} J_{il} - \del_{jl} J_{ik} - \del_{ik} J_{jl} + \del_{il} J_{jk}.
\label{1.10}
\ee
Elements of the subgroup act only on spatial coordinates $x^i$ and on \emph{spatial} components of tensor fields, \emph{e.g.} $A_i$ for a vector field.

Although we have presented some explicit examples of realisations of the algebra of spin operators on fields with a different
number of spacetime components (scalars, vectors, rank-two tensors), the question arises as to whether these 
representations of the Poincar\'{e} algebra are irreducible. One should then ask how to 
classify and realise irreducible representations of the algebra. This is done as usual by 
constructing Casimir invariants, \emph{i.e.} operators which commute with any other Lie algebra element, and studying their eigenvalue spectrum. 

We start our quest to classify representations of the Poincar\'{e} algebra by first noticing that there is a straightforward invariant arising from the nilpotent part of the algebra
\be
P^2 = P^{\mu} P_{\mu}.
\label{1.11}
\ee
This object commutes with all elements of the Poincar\'e algebra (\ref{1.1}) and is therefore a Casimir invariant. It is a hermitean operator 
with a continuous eigenvalue spectrum, which can be identified with the real line. Indeed, the eigenfunctions of $P_{\mu}$ 
are plane waves with a continuous imaginary spectrum
\be
P_{\mu}\, e^{ik \cdot x} = i k_{\mu} e^{ik \cdot x} \hs{1} \Rightarrow \hs{1} 
P^2 e^{ik\cdot x} = - k^2 e^{i k \cdot x}, 
\label{1.14}
\ee
where 
\be
- k^2 = - k^{\mu} k_{\mu} = k_0^2 - \bfk^2.
\label{1.15}
\ee
The eigenvalue spectrum $-k^2$ of $P^2$ can therefore be split into three different regions. 
First, for the negative part of the real eigenvalue spectrum, we have
\be 
- k^2 = m^2 > 0,
\label{1.16}
\ee
for some real number $m$. This part of the spectrum corresponds to time-like realisations. 
Eigenvectors in this region can be transformed by a Lorentz transformation to a coordinate 
frame in which
\be
k^{\mu} = (m, 0, 0, 0).
\label{1.17}
\ee
We call the corresponding coordinate frame the {\em rest frame}. Next, the zero eigenvalue of
$P^2$ corresponds to light-like (null) vectors: 
\be
- k^2 = 0  \Longrightarrow k_0 = \pm \sqrt{\bfk^2}.
\label{1.18}
\ee
In this case, we can find a Lorentz transformation to a coordinate frame such that
\be
k^{\mu} = (\og, 0, 0, \pm\, \og),
\label{1.19}
\ee
for some real number $\og$. Finally, there is the positive part of the spectrum of $P^2$ 
corresponding to space-like vectors 
\be
- k^2 = -\kg^2 < 0,
\label{1.20}
\ee
for some real number $\kg$. In this case a Lorentz transformation can bring $k^{\mu}$ 
to the form 
\be 
 k^{\mu} = (0,0,0,\kg).
\label{1.21}
\ee
We call such realisations {\em tachyonic}; they are generally considered unphysical and therefore are not used
to describe the physical states of classical or quantum fields. In the following we will only be concerned with time-like realisations of the Lorentz algebra. The photon is a particular light-like realisation, characterised by two possible polarisations (helicities). A complete study of light-like realisations is beyond the purpose of these notes; the interested reader is referred to the work of Wigner \cite{Wigner} and to Weinberg's book (Ref.~\cite{Weinberg:1995mt}, pages 69-72). In fact, in Sec.~\ref{Sec 2} and Sec.~\ref{linearised GR} we will recover the dynamics of the photon (graviton) as the limit of a massive vector (tensor) theory, paying particular attention to the extra polarisation modes which must become non-dynamical in the limit.

Another quadratic Casimir operator of the Poincar\'{e} algebra is the total spin squared:
\be\label{eq:SpinSquared}
\Sg^2 \equiv \frac{1}{2}\, \Sg^{\mu\nu} \Sg_{\mu\nu}.
\ee
It is trivial to check that it commutes with the translation operators $P_{\mu}$, the orbital Lorentz transformations
$M_{\mu\nu}$, and also with spin operators $\Sg_{\mu\nu}$. For a vector field $A_{\ag}$, we find 
\be
\lh \Sg^2 \rh_{\ag}^{\phant\bg} = - 3\, \del_{\ag}^{\phant\bg}, 
\label{1.23}
\ee
while for a rank-two tensor field $A_{\ag\bg}$ we have 
\be
\lh \Sg^2 \rh_{\ag\bg}^{\phant\phant\gam\del} = - 6 \del_{\ag}^{\phant\gam}\, \del_{\bg}^{\phant\del} 
  - 2\del_{\ag}^{\phant\del}\, \del_{\bg}^{\phant\gam} + 2\, \eta_{\ag\bg}\, \eta^{\gam \del}.
\label{1.24}
\ee
However, this result does not provide a complete classification and implies nothing about (ir)reducibility. 
A more systematic approach is possible by splitting the spin squared into two terms, each of 
which represents a Casimir invariant by itself, of order four in the generators. These invariants are
\be
\ba{lll}
W^2 & \equiv & \dsp{ \frac{1}{2}\, P^2\, \Sg_{\mu\nu} \lh \eta^{\mu\kg} - \frac{P^{\mu} P^{\kg}}{P^2} \rh 
 \lh \eta^{\nu\lb} - \frac{P^{\nu} P^{\lb}}{P^2} \rh \Sg_{\kg\lb}, }\\
 & & \\
Z^2 & \equiv & \dsp{ P^{\mu} \Sg_{\mu\lb} P^{\nu} \Sg_{\nu}^{\phant\lb}. }
\ea
\label{1.25}
\ee
The first invariant can be easily recognised as the square of the transverse part of the spin tensor.
On the face of it, there is
a singularity at $P^2 = 0$. In fact, this is cancelled as a result of the double contraction of a spin operator with two 
translation operators, due to the antisymmetry of $\Sg_{\mu\nu}$:
\[
P^{\mu} P^{\nu} \Sg_{\mu\nu} = 0.
\]
The second invariant is the Lorentz norm of the longitudinal component of the spin, given by
\be
Z_{\mu} = P^{\nu} \Sg_{\mu\nu}.
\ee
Note that we can define a pseudovector (adjoint of a three-form)
\be
W^{\mu} = \frac{1}{3!}\, \ve^{\mu\nu\kg\lb}\, W_{\nu\kg\lb}, \hs{1}
W_{\nu\kg\lb} = J_{\kg\lb} P_{\nu} + J_{\lb\nu} P_{\kg} + J_{\nu\kg} P_{\lb} =
\Sg_{\kg\lb} P_{\nu} + \Sg_{\lb\nu} P_{\kg} + \Sg_{\nu\kg} P_{\lb},
\label{1.27}
\ee
such that $W^2 = W_{\mu} W^{\mu}$.
This is the well-known Pauli-Ljubanski vector.
We have the decomposition 
\be 
W^2 = \frac{1}{2}\, P^2 \Sg^{\mu\nu} \Sg_{\mu\nu} - P^{\mu} \Sg_{\mu\lb} P^{\nu} \Sg_{\nu}^{\phant\lb}
\hs{1}\label{1.26}
\ee
which, using the definition Eq.~(\ref{eq:SpinSquared}), can be rewritten as
\be
P^2 \Sg^2 = W^2 + Z^2.
\ee
The last formula in (\ref{1.26}) will be of central importance in order to make a systematic classification of the Lie algebra representations.

Now we study the eigenvalue spectrum of the three Casimir invariants:
\[
P^2, \hs{1} Z^2, \hs{1} W^2.
\]
We have already discussed the spectrum of $P^2$. Let us now consider the eigenvalue problems
\be
Z^2\, F = \lb F, \hs{2} W^2\, F = \kg F.
\label{1.28}
\ee
First we consider a vector field $A_{\ag}$. We have:
\begin{align}
\left( Z^2\, A \right)_{\ag} &= - P^2 A_{\ag} - 2 P_{\ag} P^{\bg} A_{\bg} = \lb A_{\ag}, \\
\left( W^2\, A \right)_{\ag} &= - 2 P^2 A_{\ag} + 2 P_{\ag} P^{\bg} A_{\bg} = \kg A_{\ag}.\label{1.29}
\end{align}
By adding these equations, we get
\be
m^2 A_{\ag} =P^2 A_{\ag} =  - \frac{1}{3} \lh \lb + \kg \rh A_{\ag}.
\label{1.30}
\ee
By contracting the equations with $P^{\ag}$, we then obtain two solutions for $m^2 > 0$:
\begin{enumerate}[label=\emph{\alph*})]
\item a transverse vector solution
\be
\lb = - m^2, \hs{1} \kg = -2 m^2, \hs{1} P^2 A_{\ag} = m^2 A_{\ag} \hs{1} \mbox{and} \hs{1} P^{\ag} A_{\ag} = 0; 
\label{1.31}
\ee
\item a longitudinal scalar solution 
\be
\lb = - 3m^2, \hs{1} \kg = 0, \hs{1} A_{\ag} = P_{\ag}\, \Fg \hs{1} \mbox{with} \hs{1} P^2 \Fg = m^2 \Fg.
\label{1.32}
\ee
\end{enumerate}
A similar procedure can be carried out for symmetric tensor fields $A_{\ag\bg}$:
\be
\ba{lll}
\left( Z^2\, A \right)_{\ag\bg} & \hs{-.5} = \hs{-.5} & 
 - 2 P^2 A_{\ag\bg} - 4 P_{\ag} P^{\gam} A_{\bg\gam} - 4 P_{\bg} P^{\gam} A_{\ag\gam}
 +2 \eta_{\ag\bg} P^{\gam} P^{\del} A_{\gam\del} +2 P_{\ag} P_{\bg} A_{\gam}^{\phant\gam} = \lb A_{\ag\bg}, \\
 & & \\
\left( W^2\, A \right)_{\ag\bg} & \hs{-.5} = \hs{-.5} & - 6 P^2 A_{\ag\bg} + 4 P_{\ag} P^{\gam} A_{\bg\gam} 
 + 4 P_{\bg} P^{\gam} A_{\ag\gam} - 2 P_{\ag} P_{\bg} A_{\gam}^{\phant\gam} \\
 & & \\
 & & +\, 2 \eta_{\ag\bg} \lh P^2 A_{\gam}^{\;\gam} - P^{\gam} P^{\del} A_{\gam\del} \rh = \kg A_{\ag\bg}. 
\label{1.33}
\ea
\ee
In this case, there are four solutions:
\begin{enumerate}[label=\emph{\alph*})]
\item a trace-like scalar
\be
\lb = 0, \hs{1} \kg = 0, \hs{1} A_{\ag\bg} = \eta_{\ag\bg}\, \Fg, \hs{1} P^2 \Fg = m^2 \Fg;
\label{1.34}
\ee
\item a  scalar mode, associated to a traceless tensor
\be
\lb = - 8 m^2, \hs{1} \kg = 0, \hs{1} A_{\ag\bg} = \lh P_{\ag} P_{\bg} - \frac{1}{4}\, \eta_{\ag\bg} P^2 \rh \Og, \hs{1}
P^2 \Og = m^2 \Og;
\label{1.35}
\ee
\item a transverse  vector, associated to a traceless tensor
\be
\lb = - 6m^2, \hs{1} \kg = -2m^2, \hs{1} A_{\ag\bg} = P_{\ag}\, \xi_{\bg} + P_{\bg}\, \xi_{\ag}, \hs{1} 
P^2 \xi_{\ag} = m^2 \xi_{\ag}, \hs{1} P^{\ag} \xi_{\ag} = 0; 
\label{1.36}
\ee
\item a transverse traceless tensor 
\be
\lb = - 2m^2, \hs{1} \kg = -6 m^2, \hs{1} P^2 A_{\ag\bg} = m^2 A_{\ag\bg}, \hs{1} P^{\bg} A_{\ag\bg} = 0, 
 \hs{1} A_{\ag}^{\phant\ag} = 0.
\label{1.37}
\ee
\end{enumerate}
Note that the trace-like scalar solution is a zero mode of the spin Casimir operator $\Sigma^2$, given in Eq.~(\ref{1.24}), while all other modes 
are non-zero-modes with $\kg + \lb = - 8 m^2$. 
From the classification given above, we see that for $m^2 > 0$
\be
- \frac{\kg}{m^2} = s (s+1),
\label{1.38}
\ee
with $s = 0$ for scalar modes, $s = 1$ for vector modes and $s = 2$ for traceless tensor modes. This is not an accident and it is in fact part of a much more general result. We can understand the relation (\ref{1.38}) by observing 
that in the rest frame, for $m^2 > 0$, we have
\be
W^2 = \frac{m^2}{2}\, \Sg_{ij}^2, \qquad \mbox{(sum over $i,j$ is implied)}.
\label{1.39}
\ee
Defining the 3-dimensional spin pseudovector 
\be\label{eq:SpinPseudovector}
\sg_i = - \frac{i}{2}\, \ve_{ijk}\, \Sg_{jk}, 
\ee
we verify that it satisfies the familiar commutation relations of the angular momentum algebra
\be
\left[ \sg_i, \sg_j \right] = i\; \ve_{ijk} \sg_k, \hs{1} W^2 = - m^2 \sg^2, \hspace{1em} \sg^2\equiv\sg_{i}\sg_{i}.
\label{1.41}
\ee
We reproduce the well-known result that on bosonic states the squared spin has the spectrum 
\be
\sg^2 = s(s+1), \hs{1} s = 0, 1, 2, ...
\label{1.42}
\ee
while the $z$-component of the spin $\sg_z$ takes $2s+1$ values in the range $(-s, -s+1, ..., +s)$. From this we can conclude that the Pauli-Ljubanski vector, \emph{i.e.} the transverse part of $\Sg_{\mu\nu}$, contains full information about the spin properties of the field considered.

The main result of this section is that irreducible representations of the Poincar\'{e} group are labelled by two real numbers, namely the spin (as measured in the rest frame) and the mass of the field. We also saw that tensor representations are reducible, \emph{i.e.} they contain invariant subspaces (corresponding in general to different spins). In the next section, when constructing the dynamics, we will take this fact into account in order to make sure that we are not letting spurious fields (\emph{i.e.} non-physical ones) propagate. By looking at Eqs.~(\ref{1.37}),~(\ref{1.38}), we observe that a symmetric and traceless rank-two tensor carries spin two. 
 
 The main purpose of the next two sections will be that of constructing a suitable action functional for a symmetric rank-two tensor, such that only the component in Eq.~(\ref{1.37}) is dynamical. Note that we keep the field massive at this stage, since this is crucial for the derivations in Sec.~(\ref{Sec 2}). The massless case will be recovered in Sec.~(\ref{Sec 3}).
 
 Before closing this section, we want to give some physical arguments that justify the identification of the mediator of the gravitational interaction in the weak field regime with a massless spin-two field (the graviton), following Feynman \cite{Feynman:1996kb}. The first observation to make is that gravity is a long range interaction, since we know that in the static case it satisfies the inverse square law. If the field had a non-vanishing mass this would entail a screening of the gravitational interaction, as in the case of the weak interaction mediated by the W$^{\pm}$ and Z bosons. Hence the field must be massless. The second observation is that gravity is universally attractive. As all fundamental interactions, it must be mediated by a boson, \emph{i.e.} a field with integer spin. It turns out that fields with odd integer spins can lead to either attraction or repulsion, as in the case of electromagnetism. Therefore only even spins are allowed. However, a scalar particle (spin zero) would not be able to predict the observed deflection of light rays by gravity and must be excluded as well for this reason. Hence, the simplest possibility we are left with is that of a spin-two field.\footnote{Nevertheless, we would like to mention that it is possible to construct field theories with spin higher than two. We refer the interested reader to \emph{e.g.} Ref.~\cite{Weinberg:1995mt}, the review Ref.~\cite{Bouatta:2004kk} and references therein.} In fact, this argument proves to be correct and the theory obtained in Sec.~\ref{Sec 2} is indeed equivalent to General Relativity in the weak field limit, as we will see later in Sec.~\ref{Sec 3}. An important remark that must be made is that, as a consequence of the universal character of the gravitational interaction, the theory describing the dynamics of the gravitational field cannot lead to linear equations of motion. In fact, as gravity couples to all forms of energies, it must also couple to the gravitational field itself. In other words, gravity gravitates. However, since our interest is limited to gravitational waves on a flat background, we will mostly ignore the intricacies arising from the non linearity of the Einstein's field equations and we will only consider the linearised theory. Nonetheless, it is still possible to obtain the nonlinear theory from the linear one by introducing self-couplings which are consistent with gauge invariance; this will be shown in Sec.~\ref{sec:SelfInteractions}.
\newpage

\section{Deriving field equations} \label{Sec 2}
We found  in the previous section that all covariant fields, with the exception of scalar fields, carry several spin 
representations. The representation theory we have discussed sufficed to discover which fields 
carry which spin, at least in the massive case. The problem we need to solve now is to show how to obtain field equations 
for the principal spin component of a field, and for the principal spin component only. In other words,
the field equations must guarantee that only the highest spin component in a certain tensor representation propagates. Furthermore, the {\em on-shell} condition (given by the Klein-Gordon equation) must hold for such component
\[
P^2 \Fg = m^2 \Fg,
\]
not as a supplementary condition but as a consequence of the equations of motion. The remaining subsidiary spin components, which are non-physical, are required to be non-dynamical; hence, they do not propagate in spacetime.

\subsection{Field equations for vector fields}
The example of the vector field is quite illuminating in this respect, as it gives an indication of the kind of mechanism we are looking for. The 
transverse vector field must satisfy the conditions in Eq.~(\ref{1.31}). Actually this is already achieved by 
writing the eigenvalue equation for $W^2$, Eq.~(\ref{1.29}), with the proper eigenvalue $\kg = - 2m^2$. This yields the Proca equation, which gives the dynamics of a massive vector field
\be
P^2 A_{\ag} - P_{\ag} P^{\bg} A_{\bg} = m^2 A_{\ag}. 
\label{eq:ProcaChapter2}
\ee
Contracting Eq.~(\ref{eq:ProcaChapter2}) with a translation operator $P^{\ag}$ we get 
\be
m^2 P^{\ag} A_{\ag} = 0.
\label{2.2}
\ee
Therefore, for $m^2 > 0$, we automatically obtain the subsidiary condition
\be
P^{\ag} A_{\ag} = 0.
\label{2.3}
\ee
Plugging it into Eq.~(\ref{eq:ProcaChapter2}) one finds the Klein-Gordon equation, which gives the dynamics
of the remaining degrees of freedom.
In other words, the equation (\ref{eq:ProcaChapter2}) implies both the Klein-Gordon equation and the
subsidiary condition. The latter demands that the vector field be transverse, hence
killing the scalar solutions of the Klein-Gordon equation.

\subsection{Field equations for symmetric tensor fields}
We could now try to do the same for the symmetric tensor field. Thus we take the eigenvalue 
equation for $W^2$, Eq. (\ref{1.33}), with eigenvalue $\kg = - 6m^2$:
\be
P^2 A_{\ag\bg} - \frac{2}{3}\, P_{\ag} P^{\gam} A_{\bg\gam} - \frac{2}{3}\, P_{\bg} P^{\gam} A_{\ag\gam}
 + \frac{1}{3}\, P_{\ag} P_{\bg} A_{\gam}^{\phant\gam} - \frac{1}{3}\, \eta_{\ag\bg} \lh P^2 A_{\gam}^{\;\gam} 
 - P^{\gam} P^{\del} A_{\gam\del} \rh = m^2 A_{\ag\bg}.
\label{2.4}
\ee
This equation does indeed guarantee that for $m^2 > 0$ the scalar components decouple:
\be
A_{\ag}^{\phant\ag} = 0, \hs{1} P^{\ag} P^{\bg} A_{\ag\bg} = 0
\label{2.5}
\ee
thus implying
\be
P^2 A_{\ag\bg} - \frac{2}{3}\, P_{\ag} P^{\gam} A_{\bg\gam} - \frac{2}{3}\, P_{\bg} P^{\gam} A_{\ag\gam} 
 = m^2 A_{\ag\bg}.
\label{2.6}
\ee
Unfortunately this equation admits two solutions: the desired traceless tensor solution, satisfying the two conditions
\be
P^{\bg} A_{\ag\bg} = 0 \hs{1}, \hs{1} P^2 A_{\ag\bg} = m^2 A_{\ag\bg},
\label{2.7}
\ee
and the spurious longitudinal vector solution
\be
A_{\ag\bg} = P_{\ag} \xi_{\bg} + P_{\bg} \xi_{\ag} \hs{1}, \hs{1} P^2 \xi_{\ag} = 3 m^2 \xi_{\ag}. 
\label{2.8}
\ee
Thus our approach, although it works in the vector case, fails with tensors; we must therefore devise some new trick. An elegant way to get around the problem is given by the so-called {\em root method} \ct{Ogievetski:1977,Berends:1979}.
We have already observed in Eqs.~(\ref{1.34}) to (\ref{1.37}) that a generic symmetric tensor field $A_{\mu\nu}$ 
contains two spin-zero degrees of freedom, given by a longitudinal and a transverse mode
\be
A^{(0) L}_{\mu\nu} = \frac{P_{\mu} P_{\nu}}{m^2}\, \Lb, \hs{1} 
A^{(0) T}_{\mu\nu} = \lh \eta_{\mu\nu} - \frac{P_{\mu} P_{\nu}}{m^2} \rh N,
\ee
as well as a vector (spin-one) mode
\be
A^{(1)}_{\mu\nu} = P_{\mu}\, \xi_{\nu} + P_{\nu}\, \xi_{\mu}, \hs{1} P_{\mu}\xi^{\mu} = 0.
\ee
The remaining traceless and transverse symmetric tensor is the actual spin-two field
\be
A^{(2)}_{\mu\nu} = \lh \eta_{\mu\ag} - \frac{P_{\mu} P_{\ag}}{P^2} \rh A^{\ag\bg} 
  \lh \eta_{\bg\nu} - \frac{P_{\bg} P_{\nu}}{P^2} \rh - \frac{1}{3} \lh \eta_{\mu\nu} - \frac{P_{\mu} P_{\nu}}{P^2} \rh
  \lh A^{\lb}_{\phant\lb} - \frac{P^{\kg} A_{\kg\lb} P^{\lb}}{P^2} \rh. 
\ee
It is possible to construct the various spin components by means of a complete set of projection operators.
We therefore define
\be
\thg_{\mu\nu} = \eta_{\mu\nu} - \frac{P_{\mu} P_{\nu}}{P^2}, \hs{1} \og_{\mu\nu} = \frac{P_{\mu} P_{\nu}}{P^2}.
\label{2.9}
\ee
They are such that
\be
\thg_{\mu\nu} + \og_{\mu\nu} = \eta_{\mu\nu}
\label{2.10}
\ee
and
\be
\theta_{\mu\lambda}\;\theta^{\lambda\nu}=\theta_{\mu}^{\phantom{a}\nu}, \hspace{1em} \theta_{\mu\lambda}\;\omega^{\lambda\nu}=0, \hspace{1em}\omega_{\mu\lambda}\;\omega^{\lambda\nu}=\omega_{\mu}^{\phantom{a}\nu}
\ee
These are the projection operators that we have already used (implictly) for the vector case in Eq.~(\ref{eq:ProcaChapter2}), when 
separating the spin-one and spin-zero states of the vector field $A_{\mu}$.
We can use $\thg_{\mu\nu}$ and $\og_{\mu\nu}$ as building blocks for projection operators for the spin states of tensors 
with a higher rank. In the case of a symmetric rank-two tensor $A_{\mu\nu}$ the projection operators are more complicated than for the 
vector; nevertheless, they can  be given a compact 
expression in terms of $\thg_{\mu\nu}$, $\og_{\mu\nu}$
\begin{align}
\Pi_{\;\mu\nu}^{(2)\,\kg\lb} &= \frac{1}{2} \lh \thg_{\mu}^{\phantom{a}\kg}\, \thg_{\nu}^{\phantom{a}\lb} + 
 \thg_{\mu}^{\phantom{a}\lb}\, \thg_{\nu}^{\phantom{a}\kg} \rh - \frac{1}{3} \thg_{\mu\nu}\, \thg^{\kg\lb}, \label{eq:ProjectorsTensor1}\\
 \Pi_{\;\mu\nu}^{(1)\,\kg\lb} &= \frac{1}{2} \lh \thg_{\mu}^{\phantom{a}\kg}\, \og_{\nu}^{\phantom{a}\lb} + \thg_{\mu}^{\phantom{a}\lb}\, \og_{\nu}^{\phantom{a}\kg}
 +  \thg_{\nu}^{\phantom{a}\kg}\, \og_{\mu}^{\phantom{a}\lb} + \thg_{\nu}^{\phantom{a}\lb}\, \og_{\mu}^{\phantom{a}\kg} \rh, \label{eq:ProjectorsTensor2}\\
 \Pi_{\phant\phant\phant\mu\nu}^{(0T)\, \kg\lb} &= \frac{1}{3}\, \thg_{\mu\nu}\, \thg^{\kg\lb}, \hs{1}
 \Pi_{\,\phantom{a}\phant\mu\nu}^{(0L) \phantom{a}\kg\lb} = \og_{\mu\nu}\, \og^{\kg\lb}. 
\label{eq:ProjectorsTensor3}
\end{align}
The labels $L$ and $T$ stand for longitudinal and transverse, respectively. The projection operators in Eqs.~(\ref{eq:ProjectorsTensor1}),~(\ref{eq:ProjectorsTensor2}),~(\ref{eq:ProjectorsTensor3}) form a complete orthonormal set:
\be
\Pi^{(A)} \cdot \Pi^{(B)} = \del^{AB}\, \Pi^{(B)}, \hs{1} \sum_A\, \Pi^{(A)} = 1,
\label{2.12}
\ee
where $A=1,\,2,\,0L,\,0T$, and the unit symbol represents the symmetric unit tensor 
\be
1 \hs{.5} \rightarrow \hs{.5} \frac{1}{2} \lh \del_{\mu}^{\;\kg}\, \del_{\nu}^{\;\lb} + \del_{\mu}^{\;\lb}\, \del_{\nu}^{\;\kg} \rh.
\label{2.13}
\ee
The symmetric tensor field can now be decomposed in spin components
\be
A_{\mu\nu} = \sum_A\, \Pi_{\phant\mu\nu}^{(A)\, \kg\lb} A_{\kg\lb} 
 = A^{(2)}_{\mu\nu} + A^{(1)}_{\mu\nu} + A^{(0T)}_{\,\mu\nu} + A^{(0L)}_{\,\mu\nu}.
\label{2.14}
\ee
We can also define two nilpotent transition operators, interpolating between the two spin-zero components:
\be
T_{\phant\phant\phant\mu\nu}^{(LT)\,\kg\lb} = \frac{1}{\sqrt{3}}\, \og_{\mu\nu}\, \thg^{\kg\lb}, \hs{1}
T_{\phant\phant\phant\mu\nu}^{(TL)\, \kg\lb} = \frac{1}{\sqrt{3}}\, \thg_{\mu\nu}\, \og^{\kg\lb}.
\label{2.15}
\ee
It is easy to check that 
\be
\ba{l}
\left[ T^{(LT)} \right]^2 = \left[ T^{(TL)} \right]^2 = 0, \\
 \\
T^{(LT)} \cdot T^{(TL)} = \Pi^{(0L)}, \hs{1} T^{(TL)} \cdot T^{(LT)} = \Pi^{(0T)},
\ea
\label{2.16}
\ee
from which we get

\be
T^{(TL)} \cdot \Pi^{(0L)} = \Pi^{(0T)} \cdot T^{(TL)} , \hs{1} T^{(LT)} \cdot \Pi^{(0T)} = \Pi^{(0L)} \cdot T^{(LT)}.
\label{2.17}
\ee
The transition operators leave the spin-one and spin-two components unaffected
\be
T^{(A)} \cdot \Pi^{(1)} = \Pi^{(1)} \cdot T^{(A)} = 0, \hs{1}
T^{(A)} \cdot \Pi^{(2)} = \Pi^{(2)} \cdot T^{(A)} = 0, \hs{1} A = \left\{ LT, TL \right\}.
\label{2.18}
\ee

After introducing all this mathematical machinery, we are now in the position of coming back to our original problem. We will show how to derive the correct field equations, implying at the same time the Klein-Gordon equation for the spin-two component and the vanishing of the spurious lower spin components
\be
P^2 A^{(2)} = m^2 A^{(2)} \hs{1} \mbox{and} \hs{1} A^{(1)} = A^{(0T)} = A^{(0L)} = 0.
\label{eq:EquationsRankTwoComponents}
\ee
Notice that, since the projection operator $\Pi^{(2)}$ has a double pole at $P^2 = 0$, the first equation 
in Eq.~(\ref{eq:EquationsRankTwoComponents}) still contains a single pole at $P^2 = 0$. We could 
multiply the equation by another $P^2$, and write 
\be
(P^2)^2 A^{(2)} = m^4 A^{(2)}, 
\label{eq:NaiveSquareRankTwo}
\ee
but it would have tachyon solutions with $m^2 < 0$. What we need instead is a {\em regular} square root of the last equation,
corresponding to positive $m^2 > 0$ only. The trick is to add a nilpotent term to the kinetic 
operator which cancels the double pole:

\be
P^2 \lh \Pi^{(2)} + \frac{2}{\sqrt{3}}\, T^{(TL)} \rh \cdot A = m^2 A^{(2)}.
\label{2.21}
\ee
In fact the double pole term in $\Pi^{(2)}$ is
\be
\frac{2}{3}\, \frac{P_{\mu} P_{\nu} P^{\kg} P^{\lb}}{(P^2)^2}, 
\label{2.22}
\ee
while the double pole term in $T^{(TL)}$ is 
\be
- \frac{1}{\sqrt{3}}\, \frac{P_{\mu} P_{\nu} P^{\kg} P^{\lb}}{(P^2)^2}.
\label{2.23}
\ee
The coefficient of $T^{(TL)}$ in Eq.~(\ref{2.21}) has been  appropriately chosen to remove the double pole term.
Therefore, after multiplication by $P^2$, the kinetic operator acting on $A$ on the l.h.s. in Eq.~(\ref{2.21}) has 
become regular. It can be explicitly checked that by applying the same regular operator again to both sides of Eq.~(\ref{2.21}) and using the 
nilpotency of $T^{(TL)}$ we get Eq.~(\ref{eq:NaiveSquareRankTwo}). Therefore Eq.~(\ref{2.21}) is indeed a regular square root of 
Eq.~(\ref{eq:NaiveSquareRankTwo}) with $m^2 > 0$ from which the tachyon solutions have been eliminated. Written out in full, the 
field equation (\ref{2.21}) reads
\be
P^2 A_{\mu\nu} - P_{\mu} P^{\lb} A_{\nu\lb} - P_{\nu} P^{\lb} A_{\mu\lb} + \eta_{\mu\nu} P^{\kg} P^{\lb} A_{\kg\lb} 
 - \frac{1}{3} \lh \eta_{\mu\nu} P^2 - P_{\mu} P_{\nu} \rh A^{\lb}_{\;\,\lb} = m^2 A_{\mu\nu}.
\label{2.24}
\ee
It is straightforward to check that it implies
\be
P^2 A_{\mu\nu} = m^2 A_{\mu\nu}, \hs{1} P^{\lb} A_{\mu\lb} = 0, \hs{1} A^{\lb}_{\;\,\lb} = 0.
\label{2.25}
\ee
Therefore the solutions of the equations of motion for $A_{\mu\nu}$ represent a massive spin-two tensor field,
with no spurious degrees of freedom.
\newpage

\section{Linearised General Relativity}\label{linearised GR} \label{Sec 3}
In this section, we construct the action functional for a massive spin-two field and use it to recover the massless case. The theory obtained in this way is shown to be equivalent to linearised General Relativity. Of course, the theory can only be valid as an approximation in a weak field regime. This is due to the fact that gravity couples to all forms of energy, hence to the gravitational field itself. In other words, the linear theory does not take into account gravitational back-reaction. However, it is remarkable that the full nonlinear theory can still be obtained from the linear one by following the Noether construction, which yields a nonlinear theory of symmetric tensor fields similar to nonlinear $\sigma$-models for scalar fields. This bottom-up approach turns out to give results perfectly equivalent to the more conventional geometric one.

\subsection{Massive tensor fields}
\label{Subsec 3.1}
Eq.~(\ref{2.24}) gives the dynamics of a freely propagating massive spin-two field. All spurious components are projected out. The aim of this section is to obtain an action functional from which the equations of motion (\ref{2.24}) can be derived. Their derivation from an action principle requires some nontrivial steps; these will be carefully
discussed in the following.

For a free massive field $A_{\mu\nu}$, the action $S[A_{\mu\nu}]$ should be a quadratic expression
\be
S[A_{\mu\nu}] = \frac{1}{2}\, \int d^4x\, A_{\mu\nu} M^{\mu\nu\kg\lb} A_{\kg\lb},
\label{3.1}
\ee
where $M^{\mu\nu\kg\lb}$ is a second order differential operator with the following symmetry properties
\begin{align}
M^{\mu\nu\kg\lb} &= M^{\kg\lb\mu\nu}  &&\mbox{(by construction of the action)}\label{eq:FirstPropertyM}\\
&=M^{\nu\mu\kg\lb}= M^{\mu\nu\lb\kg} &&\mbox{(because the field $A_{\mu\nu}$ is symmetric)}. 
\end{align}
Stationarity of the action $S[A_{\mu\nu}]$ under arbitrary variations of the field $A_{\mu\nu}$ leads to the equations of motion\footnote{In this discussion we neglect the role played by boundary terms. This will be justified a posteriori by the observation that the theory we constructed is equivalent to Einstein's theory linearised around a flat background spacetime.}:
\be
M^{\mu\nu\kg\lb} A_{\kg\lb} = 0.
\label{3.3}
\ee
It is easy to check that the field equation (\ref{2.24}) is of the same form of \eqref{3.3} with the differential operator given by
\begin{align} \nonumber
 M^{\mu\nu\kg\lb} &= \frac{1}{2} \lh \eta^{\mu\kg} \eta^{\nu\lb} + \eta^{\mu\lb} \eta^{\nu\kg} \rh (P^{2}-m^{2}) - 
 \frac{1}{2} \lh \eta^{\nu\lb} P^{\kg} + \eta^{\nu\kg} P^{\lb} \rh P^{\mu} + \\
 \qquad &-\frac{1}{2} \lh \eta^{\mu\lb}  P^{\kg} +  \eta^{\mu\kg} P^{\lb} \rh P^{\nu} + \eta^{\mu\nu} P^{\kg} P^{\lb} - \frac{1}{3} \lh \eta^{\mu\nu} P^2 - P^{\mu} P^{\nu} \rh \eta^{\kg\lb}. 
\end{align}
It is easy immediate to see that $M^{\mu\nu\kg\lb} \neq M^{\kg\lb\mu\nu}$, so that 
Eq.~(\ref{eq:FirstPropertyM}) is not satisfied.
We can restore the symmetry of $M^{\mu\nu\kg\lb}$ by means of a field redefinition. We introduce a new symmetric tensor field $h_{\mu\nu}$ related to $A_{\mu\nu}$ as follows
\be
h_{\mu\nu} = A_{\mu\nu} - \frac{1}{3}\, \eta_{\mu\nu}\, A^{\lb}_{\;\,\lb},
\ee
which implies
\be
A_{\mu\nu} = h_{\mu\nu} - \eta_{\mu\nu}\, h^{\lb}_{\;\,\lb}.
\label{3.6}
\ee
Indeed, in terms of the new field $h_{\mu\nu}$, the field equation (\ref{2.24}) reads
\be \label{3.7}
(P^2 - m^{2}) h_{\mu\nu} - P_{\mu} P^{\lb} h_{\nu\lb} - P_{\nu} P^{\lb} h_{\mu\lb} + \eta_{\mu\nu} P^{\kg} P^{\lb} h_{\kg\lb}- \lh \eta_{\mu\nu} (P^2 - m^{2}) - P_{\mu} P_{\nu} \rh h^{\lb}_{\;\,\lb} =0,
\ee
which can be written equivalently as 
\be
\Og^{\mu\nu\kg\lb}\, h_{\kg\lb} = 0, 
\label{3.8}
\ee
where
\begin{align} \nonumber
\Og^{\mu\nu\kg\lb} & =   \frac{1}{2} \lh \eta^{\mu\kg} \eta^{\nu\lb} + \eta^{\mu\lb} \eta^{\nu\kg} \rh (P^2 - m^{2}) -\frac{1}{2} \lh  \eta^{\nu\lb} P^{\kg} +\eta^{\nu\kg} P^{\lb} \rh P^{\mu}+\\
 & - \frac{1}{2} \lh \eta^{\mu\lb} P^{\kg} + \eta^{\mu\kg} P^{\lb} \rh P^{\nu} + \eta^{\mu\nu} P^{\kg} P^{\lb}  -\eta^{\kg\lb}\lh \eta^{\mu\nu} (P^2 - m^{2}) - P^{\mu} P^{\nu}\rh.\label{eq:OmegaFreeTheory}
\end{align}
As the tensor $\Og$ is symmetric under the exchange of the first and second pair of indices, this equation of motion can indeed be derived from the action
\be
S[h_{\mu\nu}] = \frac{1}{2}\, \int d^4x\, h_{\mu\nu}\, \Og^{\mu\nu\kg\lb}\, h_{\kg\lb}.
\label{3.10}
\ee
In standard notation, by replacing $P_{\mu} \rightarrow  \partial_{\mu}$, the action $S[h_{\mu\nu}]$ takes the form
\be
\ba{lll}
S[h_{\mu\nu}] & = & -\dsp{ \frac{1}{2}\, \int d^4x \left[ \der^{\lb} h^{\mu\nu} \der_{\lb} h_{\mu\nu} - 
 2 \der_{\mu} h^{\mu\lb} \der^{\nu} h_{\nu\lb} + 2 \der_{\mu} h^{\lb}_{\;\,\lb} \der_{\nu} h^{\nu\mu} 
 - \der^{\lb} h^{\mu}_{\;\,\mu} \der_{\lb} h^{\nu}_{\;\,\nu} \rd }+\\ 
 & & \\
 & & \dsp{ \hs{4.2} \ld +\, m^2 \lh h^{\mu\nu} h_{\mu\nu} - (h^{\lb}_{\;\,\lb})^2 \rh \right]. } 
\ea 
\label{3.11}
\ee
The action $S[h_{\mu\nu}]$ is known as the {\em Fierz-Pauli} action.
The price we have paid for this construction is that the mass term is no longer of the standard form, but involves a
correction with a trace term. The physical content of the theory has  of course not changed; indeed 
contracting Eq.\ (\ref{3.7}) with the momentum operator $P^{\nu}$ one gets
\be
m^2 \lh P^{\nu} h_{\mu\nu} - P_{\mu} h^{\nu}_{\;\,\nu} \rh = 0,\label{3.12}
\ee
which, for $m^2\neq0$, implies
\be \label{3.12b}
P^{\nu} h_{\mu\nu} = P_{\mu} h^{\nu}_{\;\,\nu}.
\ee
Reinserting this result into the field equation (\ref{3.7}), the latter reads 
\be
P^2 h_{\mu\nu} - P_{\mu} P_{\nu} h^{\lb}_{\;\,\lb} = m^2 \lh h_{\mu\nu} - \eta_{\mu\nu} h^{\lb}_{\;\,\lb} \rh.
\label{3.13}
\ee
Taking the trace of this equation we get the tracelessness condition $h^{\lb}_{\;\,\lb} = 0$, which together with Eq. (\ref{3.12b}) implies that the field $h_{\mu\nu}$ is transverse, \emph{i.e.} $P^{\nu} h_{\mu\nu} = 0$.
This proves that the physical solutions are transverse and traceless. Moreover, we can see that Eq. (\ref{3.13}) simply reduces to the mass shell condition once these conditions are imposed
\be
P^2 h_{\mu\nu} = m^2 h_{\mu\nu}.
\label{3.15}
\ee
In the above derivations the non vanishing mass of the field $h_{\mu\nu}$ is crucial. In fact, it allowed us to prove that the field is transverse and traceless as a consequence of the equations of motion. The massless theory, that we are now going to focus on, can be recovered from the action of the massive theory for $m^2 = 0$. In the next subsection, we shall show that the massless action reveals a symmetry that the massive theory does not have. It is a local symmetry, analogous to the gauge symmetry of electromagnetism and Yang-Mills theories. In complete analogy with those theories, field configurations connected by a gauge transformation must be identified and regarded as completely equivalent from a physical point of view.

Before discussing the massless case, we mention an interesting property of the massless limit of massive gravity in the presence of external sources: the so-called vDVZ discontinuity (see \emph{e.g.} Ref.~\cite{Hinterbichler:2011tt}). From the discussion above it follows that a massive graviton has five degrees of freedom. In fact, $h_{\mu\nu}$ is a symmetric tensor on a four dimensional spacetime; hence it has ten independent components. Since it is transverse and traceless, we have five constraints making five of its components non-dynamical. In the massless limit, the remaining degrees of freedom are decomposed into two helicity states of the massless graviton, two helicity states of a massless vector and a massless scalar. There are no issues when considering the propagation of a free field. However, the situation is much more subtle when the field is coupled to sources. In fact, it turns out that the scalar couples to the trace of the stress-energy tensor and, therefore, it survives in the limit $m^2\to0$. The presence of the massless scalar (sometimes called longitudinal graviton) is responsible of a discontinuity in the degrees of freedom between the massive and the massless theory and has consequences on the physical content of the theory, \emph{e.g.} it gives wrong predictions for the bending of light rays. Nevertheless, by introducing new fields and gauge symmetries into massive gravity, it is possible to recover the correct massless limit. There is a well-known procedure which is used to this end, known as the St\"{u}ckelberg trick. For a detailed discussion of this issue and other aspects of massive gravity, we refer the interested reader to Ref.~\cite{Hinterbichler:2011tt}. As a side remark, the vDVZ disappears if one allows for a non-vanishing cosmological constant $\Lambda$ and then lets $m\to0$ before taking the limit $\Lambda\to0$ (see Refs.~\cite{Higuchi:1986py,Higuchi:1989gz} for de Sitter spacetime $\Lambda>0$ and Ref.~\cite{Porrati:2000cp} for anti-de Sitter spacetime $\Lambda<0$).

The vDVZ discontinuity, which shows that the massless limit of massive tensor fields and massless gravity are two distinct physical theories, does not pose any problems for our construction. In fact, our motivation for introducing a massive theory of the graviton in first place, was to have a clear procedure to obtain the mathematical structure of the action functional. From this point of view, the free massless case is obtained by formally setting the mass parameter to zero.

\subsection{Free massless tensor fields}
The action (\ref{3.11}) for $m^2 = 0$ then reduces to 
\be
S_{0}[h_{\mu\nu}] = -\frac{1}{2}\, \int d^4x \left[ \der^{\lb} h^{\mu\nu} \der_{\lb} h_{\mu\nu} - 
 2 \der_{\mu} h^{\mu\lb} \der^{\nu} h_{\nu\lb} + 2 \der_{\mu} h^{\lb}_{\;\,\lb} \der_{\nu} h^{\nu\mu} 
 - \der^{\lb} h^{\mu}_{\;\,\mu} \der_{\lb} h^{\nu}_{\;\,\nu}  \right]. 
\label{3.16}
\ee
We used the notation $S_0$ for the action of the free theory to distinguish it from the interacting one, which will be discussed later in Section \ref{sec:SelfInteractions}. The equations of motion for the massless field $h_{\mu\nu}$ are given by
\be
\Box h_{\mu\nu} - \der_{\mu} \der^{\lb} h_{\nu\lb} - \der_{\nu} \der^{\lb} h_{\mu\lb} + \der_{\mu} \der_{\nu} h^{\lb}_{\;\,\lb}
 - \eta_{\mu\nu} \lh \Box h^{\lb}_{\;\,\lb} - \der^{\kg} \der^{\lb} h_{\kg\lb} \rh = 0,
\label{3.17}
\ee
where the d'Alembertian operator is defined as $\Box \equiv \eta_{\mu\nu}\partial^{\mu}\partial^{\nu}= P^{2}$.
By taking the trace of this equation, we find that
\be
\Box h^{\lb}_{\;\,\lb} - \der^{\kg} \der^{\lb} h_{\kg\lb} = 0.
\label{3.18}
\ee
Thus, Eq.\ (\ref{3.17}) can be simplified to 
\be
\Box h_{\mu\nu} - \der_{\mu} \der^{\lb} h_{\nu\lb} - \der_{\nu} \der^{\lb} h_{\mu\lb} + \der_{\mu} \der_{\nu} h^{\lb}_{\;\,\lb} = 0.
\label{3.19}
\ee
However, contraction of Eq.~(\ref{3.17}) with $P^{\nu} =\der^{\nu}$ leads to an identity, while a similar contraction of 
Eq.~(\ref{3.19}) leads back to Eq.~(\ref{3.18}). Hence, such contractions do not lead to any new constraints on $h_{\mu\nu}$.
The reason for this can be seen after a closer inspection of the action $S_0[h_{\mu\nu}]$ and the field equations (\ref{3.19}).
Both are invariant under field transformations of the form\footnote{The action is invariant up to boundary terms.}
\be
h_{\mu\nu} \hs{0.5} \rightarrow \hs{0.5} h'_{\mu\nu} = h_{\mu\nu} + \der_{\mu} \xi_{\nu} + \der_{\nu} \xi_{\mu},
\label{3.20}
\ee
where the parameters $\xi_{\mu}(x)$ are arbitrary differentiable functions of the spacetime coordinates $x^{\mu}$. Therefore 
the transformations (\ref{3.20}) are recognised as {\em local gauge transformations}; physical configurations of the field are defined as solutions to the field  equations modulo such gauge transformations. To get a unique solution in any equivalence class, we 
may impose some extra conditions, a procedure known as {\em gauge fixing}. For the massless spin-two field $h_{\mu\nu}$, a 
convenient choice is represented by the de Donder gauge
\be
\der^{\nu} h_{\mu\nu} = \frac{1}{2}\, \der_{\mu} h^{\nu}_{\;\,\nu}.
\label{3.21}
\ee
Imposing this condition, sometimes also called harmonic gauge, the field equation (\ref{3.19}) reduces to the massless Klein-Gordon equation
\be
\Box\, h_{\mu\nu} = 0. 
\label{3.22}
\ee
This is an equation for waves travelling at the speed of light. By combining the harmonic gauge condition with the 
constraint (\ref{3.18}) one also gets
\be
\Box\, h^{\lb}_{\;\,\lb} = 0, \hs{1} \der^{\kg} \der^{\lb} h_{\kg\lb} = 0.
\label{3.23}
\ee
However, it is important to observe that the harmonic gauge condition (\ref{3.21}) does not completely eliminate the freedom to perform
gauge transformations. In fact, gauge transformations Eq.~(\ref{3.20}) with parameters satisfying the condition
\be
\Box\, \xi_{\mu} = 0
\label{3.24}
\ee
preserve the de Donder gauge and can be used to impose further restrictions on the field. 
For instance, by choosing 
\be
\der_{\mu} \xi^{\mu} = - \frac{1}{2}\, h^{\mu}_{\;\,\mu},
\label{3.25}
\ee
the trace of the new field $h'_{\mu\nu}$ in Eq.~\eqref{3.20} can be set to be zero:
\be
h^{\prime\,\mu}_{\;\;\mu} = h^{\mu}_{\;\,\mu} + 2\, \der_{\mu} \xi^{\mu} = 0.
\label{3.26}
\ee
The de Donder gauge is of course preserved and reduces to a transversality condition for the transformed field
\be
\der^{\nu} h'_{\mu\nu} = 0.
\ee
Hence, the traceless and transversality conditions are a result of a gauge fixing procedure in the massless case, whereas for the massive field they are a direct consequence of the dynamics itself, as we have shown above.

\subsection{Coupling to external sources}
In this subsection, we introduce external sources coupled to the gravitational field. This is needed in order to describe the interactions of the gravitational field with mechanical systems; in fact, we know that gravity does not simply propagate in spacetime, but couples to all sources of energy and momentum. In particular, the introduction of sources will allow us to study the emission of gravitational waves and further understand their properties, which is done in Secs.~\ref{Sec 5},~\ref{Sec 6}.
Coupling to matter is easily implemented by adding to the action $S_{0}[h_{\mu\nu}]$ in Eq.~\eqref{3.16} a new term with the coupling  $\kappa\, h_{\mu\nu}T^{\mu\nu}$, where $T_{\mu\nu}$ is the (symmetric) stress-energy tensor, which encodes the sources, and $\kappa$ is the strength of the coupling to the sources.
It is important to notice that we must have $\der^{\mu}T_{\mu\nu}=0$ in order to preserve the gauge invariance of the theory. This result is straightforward and can be obtained after performing a gauge transformation and integrating by parts.
Hence the new action is given by
\be
S[h_{\mu\nu}]=S_{0}[h_{\mu\nu}] + \kappa \int d^{4}x\; h_{\mu\nu}T^{\mu\nu},
\ee
The equations of motion \eqref{3.17} read:
\be\label{eq:eomMasslessWithSources}
\Box h_{\mu\nu} - \der_{\mu} \der^{\lb} h_{\nu\lb} - \der_{\nu} \der^{\lb} h_{\mu\lb} + \der_{\mu} \der_{\nu} h^{\lb}_{\;\,\lb}
 - \eta_{\mu\nu} \lh \Box h^{\lb}_{\;\,\lb} - \der^{\kg} \der^{\lb} h_{\kg\lb} \rh = - \kappa T_{\mu\nu}.
\ee
Tracing Eq.~(\ref{eq:eomMasslessWithSources}), one gets
\be
\Box h^{\mu}_{\;\,\mu}-\der^{\mu} \der^{\lb} h_{\mu\lb}= \frac{\kappa}{2}T^{\mu}_{\;\,\mu}.
\ee
Thus the equations of motion can also be written as:
\be
\Box h_{\mu\nu} - \der_{\mu} \der^{\lb} h_{\nu\lb} - \der_{\nu} \der^{\lb} h_{\mu\lb} + \der_{\mu} \der_{\nu} h^{\lb}_{\;\,\lb}
 = -\kappa \Bigl(T_{\mu\nu}-\frac{1}{2}\eta_{\mu\nu} T^{\lb}_{\;\,\lb}\Bigr).
\label{3.27}
\ee
As a remark, we observe that by taking the divergence of the equations of motion we can reobtain $\der^{\mu}T_{\mu\nu}=0$, \emph{i.e.}, the energy-momentum conservation law is automatically implemented.

\subsection{Self-interactions}\label{sec:SelfInteractions}
So far, we have studied a theory described by the action \eqref{3.16}, namely the free theory of a massless spin-two field which is also traceless and transverse.
In addition, we have shown that we can add a coupling term to the free action in order to take into account physical sources.
We now want to generalise the free theory and consider self-interactions (\emph{i.e.}, nonlinear terms in the equations of motion) of the massless spin-two field $h_{\mu\nu}$.
We start by observing that the action of the free theory $S_0[h_{\mu\nu}]$ in \eqref{3.16} has the following form 
\be
S_0[h_{\mu\nu}] = - \frac{1}{2}\, \int d^4x\, \der_{\rg} h_{\mu\nu} K_0^{\rg\mu\nu\sg\kg\lb}\der_{\sg} h_{\kg\lb}, 
\label{2.1}
\ee
where the kernel $K_0^{\rg\mu\kg\sg\nu\lb} = K_0^{\sg\kg\lb\rg\mu\nu}$ is a constant Lorentz tensor, explicitly given by (cf. Eqs. \eqref{eq:OmegaFreeTheory}, \eqref{3.10})
\be
\ba{lll}
K_0^{\rg\mu\nu\sg\kg\lb} & = &
 \frac{1}{2}\, \eta^{\rg\sg} \lh \eta^{\mu\kg} \eta^{\nu\lb} + \eta^{\mu\lb} \eta^{\nu\kg} \rh + 
 \frac{1}{2}\, \eta^{\mu\nu} \lh \eta^{\rg\kg} \eta^{\sg\lb} + \eta^{\rg\lb} \eta^{\sg\kg} \rh \\
 & & \\
 & &  +\, \frac{1}{2}\, \eta^{\kg\lb} \lh \eta^{\rg\mu} \eta^{\sg\nu} + \eta^{\rg\nu} \eta^{\sg\mu} \rh 
 - \eta^{\rg\sg} \eta^{\mu\nu} \eta^{\kg\lb} \\
 & & \\
 & & -\, \frac{1}{2} \lh \eta^{\rg\mu} \eta^{\sg\kg} \eta^{\nu\lb} + \eta^{\rg\nu} \eta^{\sg\lb} \eta^{\mu\kg} 
 + \eta^{\rg\nu} \eta^{\sg\kg} \eta^{\mu\lb} + \eta^{\rg\mu} \eta^{\sg\lb} \eta^{\nu\kg} \rh. 
\ea
\label{2.2}
\ee
The strategy for introducing self-interactions consists in modifying this expression for the kinetic term
by including a field-dependent correction term
\be
K^{\rg\sg\mu\nu\kg\lb} = K_0^{\rg\sg\mu\nu\kg\lb} + \Del K^{\rg\sg\mu\nu\kg\lb}[h],
\label{2.3}
\ee
where the new term $\Del K[h]$ can be expanded as a power series in the field $h_{\mu\nu}$
\be
\Del K^{\rg\sg\mu\nu\kg\lb}[h] = \sum_{n \geq 1}\kg^n K_{(n)}^{\rg\sg\mu\nu\kg\lb|\ag_1\bg_1 ... \ag_n\bg_n} 
 h_{\ag_1 \bg_1} ... h_{\ag_n \bg_n}.
\label{2.4}
\ee
This procedure is reminiscent of the construction of nonlinear $\sg$-models.
However, it is desirable to retain gauge invariance while modifying the theory, so as to leave the number of physical degrees 
of freedom unchanged and equal to that of the linear theory. Therefore, we require that the additional term in the kernel 
$K^{\rg\sg\mu\nu\kg\lb}$ is consistent with a modified form of the infinitesimal gauge transformations, which were
introduced earlier in Eq.~\eqref{3.20}
\begin{align}
\del_{\xi} h_{\mu\nu} &= \der_{\mu} \xi_{\nu} + \der_{\nu} \xi_{\mu} +
 \kg\, \xi^{\lb}\, G_{\lb\mu\nu}^{\kg\rg\sg}[h]\, \der_{\kg} h_{\rg\sg},\label{eq:ModifiedDiffeo}\\
G_{\mu\nu\lb}^{\kg\rg\sg}[h] &= G_{(0)\,\lb\mu\nu}^{\kg\rg\sg} + \sum_{n \geq 1} 
 \kg^n\, G_{(n)\,\lb\mu\nu}^{\kg\rg\sg|\ag_1\bg_1...\ag_n\bg_n} h_{\ag_1\bg_1} ... h_{\ag_n\bg_n}.
\label{2.5}
\end{align}
The requirement of modified gauge invariance actually strongly restricts the form of the functions 
$\Del K[h]$ and $G[h]$.
A first crucial observation is that the action is invariant under infinitesimal 
transformations $\del_{\xi} h_{\mu\nu}$. We require that \emph{all} symmetries are of this type.
Therefore, gauge transformations must close an algebra, \emph{i.e.}, the commutator of any two
gauge transformations is also a gauge transformation. As such, it leaves the action unchanged. For definiteness,
considering two parameters $\xi_1$, $\xi_2$, one has
\be
\del_{\xi_1} S=\del_{\xi_2} S=0,
\ee
which imply
\be
\left[ \del_{\xi_2}, \del_{\xi_1} \right] S = 0.
\ee
Since we are assuming that there are no other symmetries but gauge transformations, we have
\be
 \left[ \del_{\xi_2}, \del_{\xi_1} \right] h_{\mu\nu} = \del_{\xi_3} h_{\mu\nu},
\label{2.7}
\ee
where $\xi_3$ is by construction bilinear in the parameters $\xi_1$ and $\xi_2$ and antisymmetric 
under their interchange. Obviously, the original linear transformations with $G[h] = 0$ satisfy this 
property in a trivial way since they are Abelian.
However, after the modification
(\ref{2.5}) this is no longer true and the requirement (\ref{2.7}) imposes a non-trivial constraint.
To cut a long argument short, we observe that
Eq.~(\ref{2.7}) is satisfied to first order in $\kg$ if Eq.~(\ref{eq:ModifiedDiffeo}) is given by
\be
\del_{\xi} h_{\mu\nu} = \der_{\mu} \xi_{\nu} + \der_{\nu} \xi_{\mu} -
 2 \kg \,\xi^{\lb} \lh \der_{\mu} h_{\nu\lb} + \der_{\nu} h_{\mu\lb} - \der_{\lb} h_{\mu\nu} \rh 
 + {\cal O}\left(\kg^2\right).
\label{2.8}
\ee 
Eq.~(\ref{2.8}) implies that, given two parameters $\xi_1$, $\xi_2$, Eq.~(\ref{2.7}) is satisfied
to first order in $\kg$ if one takes
\be
\xi_{3\,\mu} = 2 \kg \lh \xi_2^{\lb} \der_{\mu} \xi_{1\,\lb} - \xi_1^{\lb} \der_{\mu} \xi_{2\,\lb} \rh.
\label{2.9}
\ee
Therefore, we find
\be
G_{(0)\,\lb\mu\nu}^{\kg\rg\sg} = - 2 \lh \del_{\mu}^{\kg} \del_{\nu}^{\rg} \del_{\lb}^{\sg} 
 + \del_{\nu}^{\kg} \del_{\mu}^{\rg} \del_{\lb}^{\sg} - \del_{\lb}^{\kg} \del_{\mu}^{\rg} \del_{\nu}^{\sg} \rh.
\label{2.10}
\ee
The next step is to look for a term $K_{(1)}^{\rg\sg\mu\nu\kg\lb|\ag\bg} h_{\ag\bg}$ which modifies the 
action so as to make it invariant under these 
extended gauge transformations, to first order in $\kg$. After some partial integrations, the result one gets is
\be
\ba{l}
K_{(1)}^{\rg\sg\mu\nu\kg\lb|\ag\bg} h_{\ag\bg} = K_0^{\sg\kg\lb\rg\mu\nu} h_{\ag}^{\;\,\ag}\, 
 + \lh h^{\mu\rg} \eta^{\nu\sg} + h^{\nu\rg} \eta^{\mu\sg} + h^{\mu\sg} \eta^{\nu\rg} 
 + h^{\nu\sg} \eta^{\mu\rg} \rh \eta^{\kg\lb}  \\
  \\
  \dsp{ \hs{3} + \lh h^{\kg\rg} \eta^{\lb\sg} + h^{\lb\rg} \eta^{\kg\sg} + h^{\kg\sg} \eta^{\lb\rg} 
 + h^{\lb\sg} \eta^{\kg\rg} \rh \eta^{\mu\nu} }\\
  \\
\dsp{ \hs{3} +\, h^{\kg\lb} \lh \eta^{\mu\rg} \eta^{\nu\sg} + \eta^{\mu\sg} \eta^{\nu\rg} \rh 
 + h^{\mu\nu} \lh \eta^{\kg\rg} \eta^{\lb\sg} + \eta^{\kg\sg} \eta^{\lb\rg} \rh  }\\
 \\
\dsp{ \hs{3} - \left[ h^{\mu\kg} \lh \eta^{\nu\sg} \eta^{\rg\lb} + \eta^{\nu\rg} \eta^{\sg\lb} \rh 
 + h^{\mu\lb} \lh \eta^{\nu\sg} \eta^{\rg\kg} + \eta^{\nu\rg} \eta^{\kg\sg} \rh \rd }\\ 
 \\
\dsp{ \hs{3} ~~+\, h^{\nu\kg} \lh \eta^{\mu\sg} \eta^{\rg\lb} + \eta^{\mu\rg} \eta^{\sg\lb} \rh 
 + h^{\nu\lb} \lh \eta^{\mu\sg} \eta^{\rg\kg} + \eta^{\mu\rg} \eta^{\kg\sg} \rh }\\
 \\
\dsp{ \hs{3} ~~+\, h^{\nu\sg} \lh \eta^{\mu\kg} \eta^{\rg\lb} + \eta^{\mu\lb} \eta^{\rg\kg} \rh 
 + h^{\nu\rg} \lh \eta^{\mu\kg} \eta^{\sg\lb} + \eta^{\mu\lb} \eta^{\kg\sg} \rh }\\
 \\
\dsp{ \hs{3} ~~+\, h^{\mu\sg} \lh \eta^{\nu\kg} \eta^{\rg\lb} + \eta^{\nu\lb} \eta^{\rg\kg} \rh 
 + h^{\mu\rg} \lh \eta^{\nu\kg} \eta^{\sg\lb} + \eta^{\nu\lb} \eta^{\kg\sg} \rh }\\
 \\
\dsp{ \hs{3} ~~+\, h^{\rg\lb} \lh \eta^{\mu\kg} \eta^{\nu\sg} + \eta^{\mu\sg} \eta^{\nu\kg} \rh 
 + h^{\rg\kg} \lh \eta^{\mu\lb} \eta^{\nu\sg} + \eta^{\mu\sg} \eta^{\nu\lb} \rh  }\\
 \\
\dsp{ \hs{3} \ld ~~+\, h^{\sg\lb} \lh \eta^{\mu\kg} \eta^{\nu\rg} + \eta^{\mu\rg} \eta^{\nu\kg} \rh 
 + h^{\sg\kg} \lh \eta^{\mu\lb} \eta^{\nu\rg} + \eta^{\mu\rg} \eta^{\nu\lb} \rh \right] }\\
 \\
\dsp{ \hs{3} + \lh h^{\mu\kg}\eta^{\nu\lb} + h^{\mu\lb} \eta^{\nu\kg} + h^{\nu\kg} \eta^{\mu\lb} 
 + h^{\nu\lb} \eta^{\mu\kg} \rh \eta^{\rg\sg} }\\
 \\
\dsp{ \hs{3} +\, h^{\rg\sg} \lh \eta^{\mu\kg} \eta^{\nu\lb} + \eta^{\mu\lb} \eta^{\nu\kg} \rh 
 - 2 \lh h^{\mu\nu} \eta^{\kg\lb} \eta^{\rg\sg}  + h^{\kg\lb} \eta^{\mu\nu} \eta^{\rg\sg} 
 + h^{\rg\sg} \eta^{\mu\nu} \eta^{\kg\lb} \rh. }
\ea
\label{2.11}
\ee
Having found the expression for the extended gauge transformations and action to first order 
in $\kg$, one can
go on and compute in an analogous fashion the corrections to order $\kg^2$. Iterating the procedure to all orders in $\kappa$
one finds an action with infinitely many terms.
However, the result is most easily written in terms of the shifted field 
\be
g_{\mu\nu} = \eta_{\mu\nu} + 2 \kg\, h_{\mu\nu}, \hs{1}
g^{\mu\nu} = \left[ \lh \eta + 2 \kg\, h \rh^{-1} \right]^{\mu\nu} = \eta^{\mu\nu} - 2 \kg\, h^{\mu\nu}
 + 4 \kg^2 h^{\mu}_{\;\,\lb} h^{\lb\nu} + ...,
\label{2.12}
\ee
where in the last expression on the r.h.s.\ all indices are still raised and lowered with the Minkowski 
metric $\eta_{\mu\nu}$ and its inverse. Then it is possible to show that, after resummation, the complete action takes the simple form 
\be
S[g_{\mu\nu}] = \frac{1}{8\kg^2}\, \int_x \sqrt{-g} \lh 2 g^{\mu\kg} g^{\nu\sg} g^{\rg\lb} - 
 2 g^{\mu\rg} g^{\nu\sg} g^{\kg\lb} - g^{\mu\kg} g^{\nu\lb} g^{\rg\sg} + g^{\mu\nu} g^{\kg\lb} g^{\rg\sg} \rh 
 \der_{\rg} g_{\mu\nu} \der_{\sg} g_{\kg\lb},
\label{2.13}
\ee
where $g\equiv \det g $ is the determinant of the shifted field
$g_{\mu\nu}$. This action is invariant (up to boundary terms) under the infinitesimal transformations 
\be
\del_{\xi} g_{\mu\nu} = \der_{\mu} \xi_{\nu} + \der_{\nu} \xi_{\mu} - 
 g^{\lb\kg} \lh \der_{\mu} g_{\kg\nu} + \der_{\nu} g_{\kg\mu} - \der_{\kg} g_{\mu\nu} \rh \xi_{\lb}.
\label{2.14}
\ee
The action (\ref{2.13}) is actually the same as Einstein's action of General Relativity, provided
we identify $g_{\mu\nu}$ with the spacetime metric. This is most easily seen by first defining the
Riemann-Christoffel symbols 
\be
\Gam_{\mu\nu}^{\lb} \equiv \frac{1}{2}\, g^{\lb\kg} \lh \der_{\mu} g_{\kg\nu} + \der_{\nu} g_{\kg\mu}
 - \der_{\kg} g_{\mu\nu} \rh, 
\label{2.15}
\ee
in terms of which the action is
\be
S[g_{\mu\nu}] = \frac{1}{2\kg^2}\, \int_x \sqrt{-g}\, g^{\mu\nu} \lh \Gam_{\mu\lb}^{\kg} \Gam_{\nu\kg}^{\lb}
  - \Gam_{\mu\nu}^{\lb} \Gam_{\lb\kg}^{\kg} \rh.
\label{2.16}
\ee
We observe that the infinitesimal transformations (\ref{2.14}) can be rewritten in the form 
\be
\del g_{\mu\nu} = \der_{\mu} \xi_{\nu} + \der_{\nu} \xi_{\mu} - 2 \Gam_{\mu\nu}^{\lb}\, \xi_{\lb}
 \equiv D_{\mu} \xi_{\nu} + D_{\nu} \xi_{\mu},
\label{eq:InfinitesimalDiffeomporphismMetric}
\ee
where we have introduced the covariant derivative of the vector parameter $\xi_{\mu}$
\be
D_{\mu} \xi_{\nu} = \der_{\mu} \xi_{\nu} - \Gam_{\mu\nu}^{\lb}\, \xi_{\lb}.
\label{eq:CovariantDerivative}
\ee
A straightforward calculation also shows that, from the definition of the Riemann-Christoffel symbols given in Eq.~(\ref{2.15}),
the covariant derivative of the metric tensor vanishes: 
\be 
D_{\lb} g_{\mu\nu} = \der_{\lb} g_{\mu\nu} - \Gam_{\lb\mu}^{\kg} g_{\kg\nu} 
 - \Gam_{\lb\nu}^{\kg} g_{\mu\kg} = 0.
\label{2.19}
\ee
Finally, after a few partial integrations the action (\ref{2.16}) can also be brought to 
the Einstein-Hilbert form:
\begin{align}
S[g_{\mu\nu}] =&  \dsp{ \frac{1}{2\kg^2}\, \int d^4x \sqrt{-g}\, g^{\mu\nu} \lh \Gam_{\mu\kg}^{\lb} \Gam_{\nu\lb}^{\kg} - 
   \Gam_{\mu\nu}^{\lb} \Gam_{\lb\kg}^{\kg} \rh }=\nonumber\\
 & \dsp{ \frac{1}{4\kg^2}\, \int d^4x \sqrt{-g}\, 
   g^{\mu\nu} \lh \der_{\lb} \Gam_{\mu\nu}^{\lb} - \der_{\mu} \Gam_{\nu\lb}^{\lb} \rh }=\nonumber\\
 & \dsp{ \frac{1}{2\kg^2}\, \int d^4x \sqrt{-g}\, g^{\mu\nu}  
   \lh \der_{\lb} \Gam_{\mu\nu}^{\lb} - \der_{\mu} \Gam_{\nu\lb}^{\lb} - \Gam_{\mu\kg}^{\lb} \Gam_{\nu\lb}^{\kg}  
   + \Gam_{\mu\nu}^{\lb} \Gam_{\lb\kg}^{\kg} \rh  }=\nonumber\\
 & \dsp{ - \frac{1}{2\kg^2}\, \int d^4x \sqrt{-g}\, R. }\label{2.20}
\end{align}
In the second step, we made use of these following identities (holding up to boundary terms)
\begin{align}
\dsp{ \frac{1}{2}\, \int \sqrt{-g}\; g^{\mu\nu} \der_{\lb} \Gam_{\mu\nu}^{\lb} }&= \dsp{ \int \sqrt{-g}\; g^{\mu\nu} \lh \Gam_{\mu\lb}^{\kg} \Gam_{\kg\nu}^{\lb} 
  - \frac{1}{2}\, \Gam_{\mu\nu}^{\lb} \Gam_{\lb\kg}^{\kg} \rh},\\
\dsp{ \int \sqrt{-g}\; g^{\mu\nu} \der_{\mu} \Gam_{\nu\lb}^{\lb} }&= \dsp{ \int \sqrt{-g}\; g^{\mu\nu} \Gam_{\mu\nu}^{\lb} \Gam_{\lb\kg}^{\kg}. }
\end{align}
In the fourth step, we used the following definition of the Riemann tensor 
\be
R_{\mu\lb\nu}^{\;\;\;\;\;\,\kg} = \der_{\mu} \Gam_{\lb\nu}^{\kg} - \der_{\lb} \Gam_{\mu\nu}^{\kg} 
 - \Gam_{\mu\nu}^{\sg} \Gam_{\lb\sg}^{\kg} + \Gam_{\lb\nu}^{\sg} \Gam_{\mu\sg}^{kg},
\ee
which gives the following expression for the Ricci tensor:
\be
R_{\mu\nu} = R_{\mu\lb\nu}^{\;\;\;\;\;\,\lb} = 
 \der_{\mu} \Gam_{\lb\nu}^{\lb} - \der_{\lb} \Gam_{\mu\nu}^{\lb} -
 \Gam_{\mu\nu}^{\sg} \Gam_{\lb\sg}^{\lb} + \Gam_{\lb\nu}^{\sg} \Gam_{\mu\sg}^{\lb}.
\ee
Alternative derivations of General Relativity from the linear theory of massless gravitons on flat spacetime can be found in Refs.~\cite{Deser:1969wk,Feynman:1996kb}, and references therein. The result was generalised to arbitrary curved backgrounds in Ref.~\cite{Deser:1987uk}.
\newpage

\section{Hamiltonian formalism } \label{Sec 4}
In the previous section, we have shown that the massless Fierz-Pauli action in Eq.~\eqref{3.16} is equivalent to the linearised Einstein-Hilbert action. In this section, we would like to bring to light the physical content of such a theory. To this end, it is convenient to introduce the Hamiltonian formalism (ADM) for linear gravity \cite{Arnowitt:1962hi}. For a canonical analysis of the massive Fierz-Pauli action, we refer the reader to Ref.~\cite{Deser:2014vqa}.

\subsection{Hamiltonian equations}
The first step is to perform a Legendre transformation of the linearised action $S_{0}[h_{\mu\nu}]$ in Eq. \eqref{3.16} only with respect to the spatial components $h_{i j}$.
We recall that the linearised action $S_{0}[h_{\mu\nu}]$ is
\begin{align}\label{eq:Fierz-Pauli}
	S_{0}[h_{\mu\nu}] &= \int \mbox{d}^{4}x\; \mathcal{L}[h_{\mu\nu}], \nonumber \\
	&=-\frac{1}{2}\, \int d^4x \left[ \der^{\lb} h^{\mu\nu} \der_{\lb} h_{\mu\nu} - 
	2 \der_{\mu} h^{\mu\lb} \der^{\nu} h_{\nu\lb} + 2 \der_{\mu} h^{\lb}_{\;\,\lb} \der_{\nu} h^{\nu\mu} 
	- \der^{\lb} h^{\mu}_{\;\,\mu} \der_{\lb} h^{\nu}_{\;\,\nu}  \right], 
\end{align}
and it is accompanied by the interaction term between the gravitational field and matter 
\be\label{eq:interaction term}
S_{int}[h_{\mu\nu}] =\kappa \int \mbox{d}^{4}x\; h^{\mu\nu}T_{\mu\nu}.
\ee
Before introducing the Hamiltonian formalism, we must pick one representative in the class of actions equivalent to Eq.~\eqref{eq:Fierz-Pauli}, \emph{i.e.} up to a surface term. We observe that it is possible to write down an action, which is equivalent to Eq.~\eqref{eq:Fierz-Pauli}, where the only velocities appearing are those of the spatial components of the metric $h_{ij}$. For the following it is important to notice that, given the conventions we adopted for the metric signature, spatial indices can be lowered or raised without any change of sign, \emph{e.g.} $\pi_{i j}= \pi^{i j}$, whereas raising or lowering a timelike index implies a change of sign.
It is useful to define the variables
\be
N\equiv-h_{00}, \hspace{1em} N_i\equiv2h_{0i},
\ee
respectively called lapse function and shift vector. It is clear from this definition and the metric signature convention we adopted that
\be
N^i=2h_0^{~ i}=N_i.
\ee
It is possible to rewrite the action given by Eq. (\ref{eq:Fierz-Pauli}) with the interaction term Eq.~(\ref{eq:interaction term})
\be
S=S_0+S_{int}
\ee
in a different form, by splitting time and space components of the tensor fields. This is consistent with the spirit of the canonical formalism, where space and time are no longer treated on the same footing. In fact, after some partial integrations one gets
\begin{align}
	&S[h_{ij},N,N_i]=\label{eq:equivalent action}\\
	&=\int \mbox{d} t \int \mbox{d}^{3}x\; \left[ \frac{1}{2}(\dot{h}_{ij})^2-\frac{1}{2}(\dot{h}_{ii})^2-\frac{1}{2}(\partial_{k}h_{ij})^2+\frac{1}{4}(\partial_{k}h_{ii})^2+\left(\partial_{i}h_{ij}-\frac{1}{2}\partial_{j}h_{ii}\right)^2+\right.\nonumber\\
	&+\left. N(\partial_{i}\partial_{j}h_{ij} -\partial_i\partial_i h_{jj})-\partial_jN_i(\dot{h}_{ij}-\delta_{ij}\dot{h}_{kk})+\frac{1}{8}(\partial_i N_j-\partial_j N_i)^2-\kappa N T_{00} -\kappa N_{i}T_{0i}+\kappa h_{ij}T_{ij}\right].\nonumber
\end{align}
Also notice that, at this stage, we only deal with spatial indices, which can all be lowered without creating sign confusion. The only fields whose velocities appear in the action are the spatial components of the metric perturbation $h_{ij}$, with the canonical momenta given by ($\mathcal{L}$  obviously stands for the integrand in Eq. (\ref{eq:equivalent action}))
\be\label{eq:relation momentum velocity}
\pi_{i j}=\frac{\partial \mathcal{L}}{\partial \dot{h}_{i j}}=\dot{h}_{i j} - \dot{h}_{l l}\delta_{i j} -\frac{1}{2} (\partial_{i}N_{j}+\partial_{j}N_{i}) + \partial_{k} N_{k}\delta_{i j}.
\ee
On the other hand, the canonically conjugate momenta to $N$ and $N_i$ vanish identically, since their time derivatives do not appear in the action. Therefore, these fields ought to be treated as Lagrange multipliers.
Inverting Eq.~(\ref{eq:relation momentum velocity}) so as to solve for the velocities $\dot{h}_{i j}$ and noticing that $\pi_{k k}=-2(\dot{h}_{k k} - \partial_{k} N_{k}$), we get
\be\label{eq:inverse relation velocity momentum}
\dot{h}_{i j}=\pi_{i j} - \frac{1}{2}\pi_{k k}\delta_{i j} +\frac{1}{2} (\partial_{i}N_{j}+\partial_{j}N_{i}).
\ee
One can now use Eq.~(\ref{eq:inverse relation velocity momentum}) to compute the canonical Hamiltonian and rewrite the action in Eq.~(\ref{eq:equivalent action}) by means of a Legendre transformation
\be\label{eq: action in canonical form}
S[h_{i j}, \pi_{i j}, N, N_i] =  \int \mbox{d}t \int \mbox{d}^{3}x\; \left(\pi_{i j}\dot{h}_{i j}-\mathcal{H}\right).
\ee
At this stage, $\pi_{ij}$ can be understood as an auxiliary field. In fact, by computing the equations of motion one finds again Eq.~(\ref{eq:relation momentum velocity}), which can be used to eliminate $\pi_{ij}$ and go back to the original action in Eq.~(\ref{eq:Fierz-Pauli}). Alternatively, one can use Eq.~(\ref{eq:inverse relation velocity momentum}) to eliminate the velocity and express the action solely in terms of phase space variables.
The Hamiltonian density is given by (cf. \emph{e.g.} \cite{Hinterbichler:2011tt})
\begin{align}\label{eq:Hamiltonian}
	\mathcal{H}&=\frac{1}{2}(\pi_{i j})^2 - \frac{1}{4}(\pi_{k k})^2 +\frac{1}{2}(\partial_{k}h_{i j})^2 - \frac{1}{4}(\partial_{j}h_{i i})^2 -\left(\partial_{i}h_{i j} -\frac{1}{2}\partial_{j}h_{ii}\right)^2-\nonumber\\
	&-N\left(\partial_{i}\partial_{j}h_{i j}-\partial_{i}\partial_{i}h_{jj}-\kappa T_{00}\right)-N_i\left(\partial_{j}\pi_{i j}-\kappa T_{0i}\right)-\kappa h_{ij}T_{ij}.
\end{align}
Variation with respect to the Lagrange multipliers yields
\be \label{constraints}
\partial_{j}\pi_{i j}=\kappa T_{0i}, \qquad \partial_{i}\partial_{j}h_{i j}-\partial_{i}\partial_{i}h_{jj}=\kappa T_{00}.
\ee
The latter are interpreted as constraint equations, respectively called (spatial) \emph{diffeorphism} and \emph{Hamiltonian constraint}. It is possible to show that the constraints are first class in the sense of Dirac, \emph{i.e.} they close a Poisson algebra \cite{Hinterbichler:2011tt}. This is a general property of gauge theories \cite{Henneaux:1992ig}.

The canonical equations of motion are obtained in the usual way, by varying the action Eq.~(\ref{eq: action in canonical form}) with respect to the canonical variables\footnote{When varying the action one should make sure that variations of the fields are contracted with symmetric tensor in order to obtain consistent dynamics. This can also be achieved by computing the equations of motion naively and then symmetrising them.} $h_{ij}$ and $\pi_{ij}$; they are
\begin{align} \label{EOM1}
	\dot{h}_{i j} &= \pi_{i j} - \frac{1}{2}\pi_{k k}\delta_{i j} +\frac{1}{2}(\partial_{i}N_j+\partial_{j}N_i),\\
	-\dot{\pi}_{i j} &=  -\partial_k\partial_k h_{i j}+\left(\partial_l\partial_l h_{kk}-\partial_l\partial_k h_{lk}\right)\delta_{ij}+\partial_i\partial_k h_{kj}+\partial_j\partial_k h_{ki}-\partial_i\partial_j h_{kk}- \nonumber\\ \label{EOM2}
	&\hs{5.3} -\partial_i\partial_j N+\partial_k\partial_k N \delta_{ij}-\kappa T_{ij},
\end{align}
while the evolution of the spatial trace of the canonical momenta is given by
\be \label{EOM3}
-\dot{\pi}_{kk}= \partial_k \partial_k h_{j j} -  \partial_k \partial_j h_{kj} + 2 \partial_k \partial_k N-\kappa T_{ii}.
\ee
Notice that while the constraints do not involve the Lagrange multipliers, the Hamiltonian equations do.
Moreover, observe that, using the second constraint, we can rewrite Eq.~\eqref{EOM3} as
\be \label{EOM4}
\dot{\pi}_{kk}= -2 \partial_k \partial_k N+\kappa (T_{00}+T_{kk}).
\ee

\subsection{Fixing the gauge: the TT gauge}
Let us now come to the local gauge transformations expressed in Eq.~\eqref{3.20} and see how they are realised in the canonical formalism. Defining the gauge parameters as $\xi_\mu=(a,\mathbf{a})$, the canonical variables transform as
\begin{align}
	h^{\prime}_{ij}&=h_{ij}+\partial_{i}a_j+\partial_{j}a_i,\label{eq:gauge transformation h spatial}\\
	\pi_{ i j}&=\pi_{ij} +2 \delta_{ij}\partial_k\partial_k a - 2\partial_{i}\partial_j a,  \label{transf pi}
\end{align}
while the Lagrange multipliers transform as
\begin{align}
	N^{\prime}&=N- 2\dot{a},\label{eq:gauge transformation h time}\\
	N^{\prime}_{i}&=N_{i}+ 2(\dot{a}_{i} +\partial_i a)\label{eq:gauge transformation h mixed}.
\end{align}
It is a convenient choice to impose the de Donder gauge condition, which is reminiscent of the Lorentz gauge used in electrodynamics,
\be\label{eq:DeDonder}
\partial^{\mu}\overline{h}_{\mu\nu}=0,
\ee
where we introduced the new tensor $\overline{h}_{\mu\nu}=h_{\mu\nu}-\frac{1}{2}\eta_{\mu\nu}h$. The advantage of this condition over other possible choiches, like \emph{e.g.} $\partial^{\mu}h_{\mu\nu}=0$, is readily seen. In fact, under a local gauge transformation, Eq.~\eqref{eq:DeDonder} reads
\be
\partial^{\mu}\overline{h}^{\;\prime}_{\mu\nu}=\partial^{\mu}\overline{h}_{\mu\nu}+\partial^{\mu}\partial_{\mu}\xi_{\nu}.
\ee
It is therefore always possible to realise the gauge condition $\partial^{\mu}\overline{h}^{\;\prime}_{\mu\nu}=0$ by choosing $\xi_{\mu}$ so as to satisfy the wave equation with a source term
\be
\partial^{\mu}\partial_{\mu}\xi_{\nu}=-\partial^{\mu}\overline{h}_{\mu\nu},
\ee
whose solution is uniquely determined once the initial data on a spacelike hypersurface is assigned. Although, this still leaves a residual gauge freedom. In fact, the de Donder gauge is still satisfied after performing a transformation with gauge parameters $\xi_\mu$ satisfying the (sourceless) wave equation.
The TT gauge is defined by the two further conditions: the traceless $h'=h'^{\mu}_{\;\;\mu}=0$ and the transverse condition $h'_{0i}=0$ or, equivalently, $N^{\prime}+h^{\prime}_{kk}=0$ and $N^{\prime}_i=0$. 
They fix the gauge completely, leaving only the physical degrees of freedom of $h_{\mu\nu}$. 
It follows from the transformation law for the metric, Eqs. (\ref{eq:gauge transformation h spatial}), (\ref{eq:gauge transformation h time}), (\ref{eq:gauge transformation h mixed}), that the above conditions can be realised on a fixed time $t=t_0$ hypersurface by requiring
\begin{align}\label{gauge TT first set}
	2(\dot{a}_{i}+\partial_{i}a)&=-N_{i},\\
	2(\dot{a}-\partial_{k}a_{k})&=h.
\end{align}
We also demand that the time derivatives of these equations vanish at $t=t_0$. Using $\partial^{\mu}\partial_{\mu}\xi_{\nu}=0$, we have
\begin{align}\label{gauge TT second set}
	2(\partial_{k}\partial_{k}a_{i}+\partial_{i}\dot{a})&=-\dot{N}_{i},\\
	2(\partial_{k}\partial_{k}a-\partial_{k}\dot{a}_k)&=\dot{h}.
\end{align}

The sets of Eqs.~(\ref{gauge TT first set}) and (\ref{gauge TT second set}) can be solved on the initial hypersurface, yielding $a|_{t_{0}}$, $\dot{a}|_{t_{0}}$, $a_{i}|_{t_{0}}$, $\dot{a}_{i}|_{t_{0}}$. Using these as initial conditions, we can solve the wave equation and determine uniquely the gauge transformation that realises the TT gauge \cite{Wald:1984rg}. The gauge fixing procedure is consistent with the dynamics, provided that there are no sources in the spacetime region we are considering, in analogy with the radiation gauge in electromagnetism.

The argument given above applies to the standard covariant formulation of General Relativity and fixes the gauge d.o.f. completely \emph{in vacuo}. What happens when we apply it to its Hamiltonian formulation? It turns out that all of the above does not change; however, we have to make sure that the dynamics of $h$ as given by the Hamiltonian equations of motion is compatible with the gauge conditions. Notice that $N_{i}$ plays no role in this discussion, since it is only a Lagrange multiplier. Furthermore, since in the TT gauge $N_{i}^{\prime}=h=0$, the de Donder condition, Eq.~(\ref{eq:DeDonder}), in the new gauge reads
\be\label{eq:DeDonderTTgaugeNprime}
\frac{\partial N^{\prime}}{\partial t}=0.
\ee
Therefore, in order for the TT gauge to be preserved over time, it is necessary that $h^{\prime}_{kk}=0$. 
With this choice, from Eq. \eqref{EOM1} we get that
\be
\dot{h}^{\prime}_{i j} = \pi^{\prime}_{i j} - \frac{1}{2}\pi^{\prime}_{kk}\delta_{i j} \Longrightarrow \dot{h}^{\prime}_{kk}=-\frac{1}{2}\pi^{\prime}_{kk}.
\ee
By restricting to $t=t_0$, we conclude that compatibility of the gauge condition with the dynamics requires imposing an initial condition on the trace of the momentum
\be
0=\left.\dot{h}^{\prime}_{kk} \right| _{t_{0}}=\left.\pi^{\prime}_{kk}\right|_{t_{0}}.
\ee
Moreover, from the Hamiltonian equations of motion we have
\be\label{eq:gauge fixing pi dot}
\ddot{h}^{\prime}_{kk}=-\frac{1}{2}\dot{\pi}^{\prime}_{kk}= \partial_i \partial_i N^{\prime}-\frac{\kappa}{2} (T_{00}+T_{kk})=0.
\ee
Consistency with Eq.~(\ref{eq:DeDonderTTgaugeNprime}) is ensured if one demands that sources vanish throughout spacetime.
We can finally conclude that after imposing the de Donder gauge, the TT gauge is fixed and is compatible with the dynamics \emph{in vacuo} provided that $h^{\prime}_{kk}=\dot{h}^{\prime}_{kk}=\pi^{\prime}_{kk}=0$ on a fixed time hypersurface $t=t_0$.

\subsection{A simple example: gravitational field of a point-like mass}
We are now going to determine the gravitational field generated by a point mass as a perturbation of the Minkowskian background. We choose an inertial frame such that the position of the mass is fixed. For convenience, the point mass is placed at the origin of our coordinate system. The only non-vanishing component of the stress-energy tensor of the particle is
\be
T_{00}=m\;\delta^{(3)}(\mathbf{r}).
\ee
As shown above, in the presence of sources \emph{it is not possible} to realise the TT gauge since $N$ cannot be made to vanish. In this case, the dynamics is entirely given by the constraint Eq.~(\ref{EOM4}), which leads to
\be\label{eq:Newton equation}
\partial_k\partial_k N=\frac{\kappa}{2} T_{00}.
\ee
Furthermore, given the isotropy of the stress-energy tensor, using the second constraint in Eq.~(\ref{constraints}) and by comparison with Eq.~(\ref{eq:Newton equation}), one gets
\be
h_{ij}=-N\delta_{ij}.
\ee
The solution is a static spacetime, with lapse function given by the Newtonian potential
\be
N= -h_{00}=-\frac{\kappa m}{8\pi r} .
\ee
The perturbed spacetime metric can be reconstructed adding the perturbation to the Minkowskian background
\be
g_{\mu\nu}=\eta_{\mu\nu}+2\kappa\; h_{\mu\nu}.
\ee
The factor $\kappa$ was inserted in order to give the perturbations canonical dimension of mass $[h_{\mu\nu}]=[M]$.
Since $\kappa^2=8\pi G$, one has
\be
\mbox{d}s^2=-\left(1-\frac{2Gm}{r}\right) dt^2+\left(1+\frac{2Gm}{r}\right)(dr^2 + r^2 d\Omega^2).
\ee
In the weak field regime, \emph{i.e.}, far from the massive point source, one can recognise this line element as the linearisation of the Schwarzschild solution in isotropic coordinates
\be
\mbox{d}s^2 = \dsp - \left( \frac{2r - Gm}{2r + Gm} \right)^2 d t^2 + 
 \left(1 + \frac{Gm}{2r} \right)^4 \left( d r^2 + r^2 d\Omega^2 \right). 
 \ee

\newpage
\section{Gravitational waves} \label{Sec 5}
This section is mainly devoted to the physical implications of the linearised equations of motion, \emph{i.e.}, what General Relativity predicts about gravitational dynamics in the weak field regime. The existence of a \emph{wave zone} in General Relativity, \emph{i.e.} a region where gravitational waves propagate freely and satisfy the superposition principle (despite the nonlinearity of the theory), was proved in Ref.~\cite{PhysRev.121.1556}. Its definition requires the background spacetime to be asymptotically flat.

\subsection{Energy flux carried by gravitational waves}
In the following, we fix the de Donder gauge with the additional conditions $h_{kk}=\pi_{kk}=N_i=0$ and impose the Hamiltonian constraint Eq.~(\ref{constraints}).
The Hamiltonian equations of motion, Eqs.~\eqref{EOM1} and \eqref{EOM2}, take the simpler form
\begin{align} \label{Ham equations}
\dot{h}_{i j} &= \pi_{i j} ,\\
-\dot{\pi}_{i j} &=  -\partial_k\partial_k h_{i j} +\partial_i\partial_k h_{kj}+\partial_j\partial_k h_{ki} -\partial_i\partial_j N -\frac{\kappa}{2}(T_{00}-T_{kk}) \delta_{ij}-\kappa T_{ij}.
\end{align}
The Hamiltonian density, Eq.~(\ref{eq:Hamiltonian}), reads
\be
\mathcal{H}=\frac{1}{2}\pi_{ij}^2+\frac{1}{2}(\partial_k h_{ij})^2-(\partial_j h_{ij})^2-\kappa h_{ij}T_{ij}.
\ee
Its time derivative is given by
\be
\frac{\partial\mathcal{H}}{\partial t}=\pi_{ij}\dot{\pi}_{ij}+\partial_k h_{ij}\partial_k \dot{h}_{ij}-2\partial_j h_{ji}\partial_k \dot{h}_{ki}-\kappa \dot{h}_{ij}T_{ij}-\kappa h_{ij}\dot{T}_{ij}.
\ee
We define the flux of energy as
\be\label{eq:GWenergyflux}
\mathcal{P}_k=\pi_{ij}\partial_{k}h_{ij}-2\pi_{ki}\partial_{j}h_{ji}.
\ee
Together, $\mathcal{H}$ and $\mathcal{P}_k$ satisfy a continuity equation with a source term
\be
\frac{\partial\mathcal{H}}{\partial t}=\partial_k \mathcal{P}_k +\pi_{ij}\left[\partial_i\partial_j N+\frac{\kappa}{2}(T_{00}-T_{kk}) \delta_{ij}\right]- kh_{ij}\dot{T}_{ij}.
\ee
In vacuo, it is possible to realise the TT gauge with the extra condition $N=0$ and the above reduces to a homogeneous continuity equation
\be\label{eq:continuity}
\frac{\partial\mathcal{H}}{\partial t}=\partial_k \mathcal{P}_k.
\ee
Eq.~\eqref{eq:continuity} thus implies that energy can be carried by the gravitational field, see also Ref.~\cite{PhysRev.121.1556}.
However, we should point out that this conclusion rests on the assumption that the background is Minkowski spacetime. 
In fact, in the full theory of General Relativity it is not possible to define a fully covariant local stress-energy 
tensor for the gravitational field, as a consequence of the strong equivalence principle\cite{Misner:1974qy}.

As shown before, in a region with no sources it is possible to realise all of the conditions that define the TT gauge. In particular, the de Donder gauge condition implies
\be
0=\partial^{\mu}\overline{h}_{\mu j}=\partial^{\mu}h_{\mu j}=-\dot{h}_{0 j}+\partial_k h_{k j}=-\frac{1}{2}\dot{N}_i+\partial_k h_{k j}.
\ee
Since in the TT gauge $N_i=0$ we end up with
\be
\partial_k h_{k j}=0,
\ee
which, together with the definition Eq.~(\ref{eq:GWenergyflux}), yields the following expression for the energy flux in the $k$-direction carried by gravitational waves in vacuo
\be\label{eq:energyFluxVacuo}
\mathcal{P}_k=\pi_{ij}\partial_{k}h_{ij}.
\ee

\subsection{Propagation of gravitational waves in Fourier space}
In the weak field limit, where our formalism for linearised gravity holds, it is convenient to study the propagation of gravitational waves using the frequency representation. Therefore, we introduce the Fourier transforms of the non vanishing metric components
\be\label{eq:transforms}
h_{ij}(t,\mathbf{r})=\int\frac{\mbox{d}\omega}{\sqrt{2\pi}}\; e^{-i\omega t}\varepsilon_{ij}(\omega,\mathbf{r}),\hspace{1em} N(t,\mathbf{r})=\int\frac{\mbox{d}\omega}{\sqrt{2\pi}}\; e^{-i\omega t}n(\omega,\mathbf{r}).
\ee
We require the polarisation tensor $\varepsilon_{ij}$ and the scalar $n$ to satisfy the following properties, which guarantee the reality of the fields
\be\label{eq:realityConditions}
\varepsilon_{ij}^{*}(\omega,\mathbf{r})=\varepsilon_{ij}(-\omega,\mathbf{r}), \hspace{1em} n^{*}(\omega,\mathbf{r})=n(-\omega,\mathbf{r}).
\ee
We also introduce the Fourier decomposition of the stress-energy tensor of the sources
\be
T_{\mu\nu}(t,\mathbf{r})=\int\frac{\mbox{d}\omega}{\sqrt{2\pi}}\;e^{-i\omega t}\tau_{\mu\nu}(\omega,\mathbf{r}).
\ee
In the following, we drop the $\omega$ and $\mathbf{r}$ dependence of $\epsilon_{ij}$, $n$  and $\tau_{ij}$ in order to make the notation lighter.
From the Hamiltonian equations of motion, Eqs.~\eqref{Ham equations}, we get
\be
 \partial_k\partial_k h_{i j} -\ddot{h}_{i j} = \partial_i\partial_k h_{kj}+\partial_j\partial_k h_{ki} -\partial_i\partial_j N -\frac{\kappa}{2}(T_{00}-T_{kk}) \delta_{ij}-\kappa T_{ij}.
\ee
In the frequency domain this looks like
\be
\left(\partial_k \partial_k +\omega^2\right)\varepsilon_{ij}=\partial_i\partial_k \varepsilon_{kj}+\partial_j\partial_k \varepsilon_{ki}-\partial_i\partial_j n -\kappa\tau_{ij}-\frac{\kappa}{2}\left(\tau_{00}-\tau_{kk}\right)\delta_{ij}\equiv \kappa m_{ij}.
\ee
The first two terms after the first equal sign can be rewritten, using Eqs.~(\ref{constraints}),~(\ref{eq:transforms}), as
\be
\partial_i\partial_k \varepsilon_{kj}+\partial_j\partial_k \varepsilon_{ki} = \frac{i \kappa}{\omega}\left(\partial_{i}\tau_{0j} + \partial_{j}\tau_{0i}\right).
\ee
At this point it is worth trying to get some insight into the physical problem at hand. We assume that the sources are localised in a region of spacetime $\Sigma$ having compact spatial slices. This means that, for any given inertial frame selected using the Minkowski background, the motion of the sources is spatially confined at all times. 
Since we know that the fundamental solution to the Helmholtz equation obeys
\be
\left(\partial_k \partial_k +\omega^2\right)\frac{e^{i\omega r}}{4\pi r}=-\delta^{(3)}(\mathbf{r}),
\ee
the solution to the wave equation is given by a convolution of the above with the source
\be\label{eq:SolutionEpsilon}
\varepsilon_{ij}(\omega, \mathbf{r}) = - \frac{\kappa}{4\pi}\int \mbox{d}^3\mathbf{r^{\prime}} \frac{e^{i\omega |\mathbf{r}-\mathbf{r^{\prime}}|}}{|\mathbf{r}-\mathbf{r^{\prime}}|} m_{ij}(\omega, \mathbf{r^{\prime}}) \equiv \frac{\kappa}{4\pi} \frac{e^{i \omega r}}{r} t_{ij}(\omega, \mathbf{r}),
\ee
where we have factored out $\frac{1}{r}$ and defined
\be \label{tij1}
t_{ij}(\omega, \mathbf{r})=-\int\mbox{d}^3\mathbf{r^{\prime}}\;\frac{r}{|\mathbf{r}-\mathbf{r^{\prime}}|}e^{i\omega(|\mathbf{r}-\mathbf{r^{\prime}}|-r)}m_{ij}(\omega,\mathbf{r^{\prime}}).
\ee
Now we look at asymptotic solutions for very large $r$. Expanding in powers of $\frac{r^{\prime}}{r}$, we get for factor of the integrand of Eq. \eqref{tij1} the following expansion
\be
\frac{r}{|\mathbf{r}-\mathbf{r^{\prime}}|}e^{i\omega(|\mathbf{r}-\mathbf{r^{\prime}}|-r)}=e^{-i\omega \mathbf{\hat{r}}\cdot\mathbf{r^{\prime}}}\left\{1+\frac{\mathbf{\hat{r}}\cdot\mathbf{r^{\prime}}}{r}+\frac{i\omega}{2r}\left[(r^{\prime})^2-(\mathbf{\hat{r}}\cdot\mathbf{r^{\prime}})^2\right]+\mathcal{O}\left(\frac{1}{r^2}\right)\right\}.
\ee
The $\mathcal{O}(1)$ terms of the integral in Eq. \eqref{tij1} are given by
\be \label{tij2}
t_{ij}(\omega, \mathbf{r})=\int\mbox{d}^3\mathbf{r^{\prime}}e^{-i\omega \mathbf{\hat{r}}\cdot\mathbf{r^{\prime}}}\left[\frac{\omega^2}{\kappa}\hat{r}_i\hat{r}_jn+\frac{1}{2}(\tau_{00}-\tau_{kk})\delta_{ij}+(\hat{r}_i\tau_{0j}+\hat{r}_j\tau_{0i})+\tau_{ij}\right]+ \mathcal{O}\left(\frac{1}{r}\right).
\ee
We can further simplify $t_{ij}(\omega, \mathbf{r})$. Indeed, from the conservation of the energy-momentum tensor, $\partial^{\mu}T_{\mu\nu}=0$, and the Hamiltonian constraint $\partial_k\partial_k N = \frac{\kappa}{2}(T_{00}+T_{kk})$, we get the following equations
\be
-i \omega \tau_{00} = \partial_i \tau_{0i}, \quad  -i \omega \tau_{0j} = \partial_i \tau_{ij},\quad \partial_i\partial_j\tau_{ij} = -\omega^2 \tau_{00}, \quad \partial_i\partial_i n=\frac{\kappa}{2}(\tau_{00}+\tau_{kk}),
\ee
which can be used to write Eq. $\eqref{tij2}$ as
\begin{align} \label{tij3}
t_{ij}(\omega, \mathbf{r})&=\int\mbox{d}^3\mathbf{r^{\prime}}e^{-i\omega \mathbf{\hat{r}}\cdot\mathbf{r^{\prime}}}\left[P_{il}\left(\tau_{lm}+ \frac{1}{2}(\hat{r}_p\tau_{pq}\hat{r}_q-\tau_{qq})\delta_{lm}\right)P_{mj}\right],\nonumber\\
&=\frac{\omega^2}{2}P_{il}M_{lm}P_{mj},
\end{align}
where we have defined the projection tensor $P_{ij} = \delta_{ij} - \hat{r}_i\hat{r}_j$, and $M_{lm}=I_{lm} +\frac{1}{2}\hat{r}_p I_{pq}\hat{r}_q\delta_{lm}$, where $I_{ij}$ is defined by
\be
\frac{\omega^2}{2}I_{ij}(\omega,\mathbf{r})=\int\mbox{d}^3\mathbf{r^{\prime}}e^{-i\omega \mathbf{\hat{r}}\cdot\mathbf{r^{\prime}}}\left(\tau_{ij}(\omega,\mathbf{r^{\prime}}) - \frac{1}{3}\tau_{kk}(\omega,\mathbf{r^{\prime}})\delta_{ij}\right).
\ee

\subsection{Energy radiated by sources}
Let us compute the energy radiated by sources contained in a compact region of space $\Sigma$, as depicted in Fig.~1. Outside $\Sigma$ the stress energy tensor is assumed to vanish, so that we are in vacuo. 
\begin{figure}[h] 
\centering
\label{sigma}
\def\svgscale{1.0}
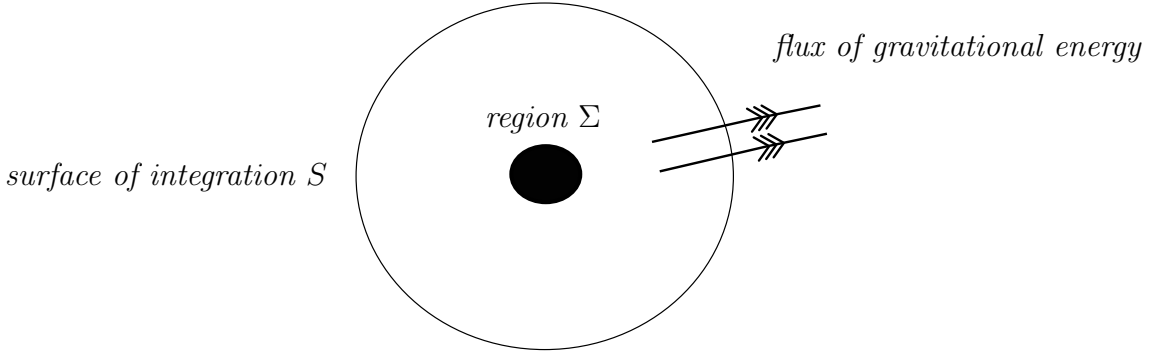
\caption{Schematic representation of a source emitting gravitational waves, enclosed by the surface of integration $S$.}
\end{figure}

We can therefore enclose such a region with a spherical surface $S$ and compute the flux of energy through $S$ in the form of gravitational waves.
\be\label{eq:EnergyRadiated}
E_{\textrm{rad}}=\int \de t\; \frac{\de E}{\de t}=\int \de t \int \de^3 \mathbf{r}\; \frac{\partial \mathcal{H}}{\partial t}=\int \de t \int \de^3 \mathbf{r}\; \partial_{k}\mathcal{P}_k=\int \de t \oiint \de\Omega\; r^2 \mathcal{P}_r.
\ee
The above was evaluated using the continuity equation Eq.~(\ref{eq:continuity}) and the divergence theorem applied to the spherical surface $S$ and its interior. Since we know that the energy flux in vacuo is given by Eq.~(\ref{eq:energyFluxVacuo}), we can compute its integral over time using the Fourier decomposition Eq.~(\ref{eq:transforms}) and the reality conditions Eq.~(\ref{eq:realityConditions}). Thus
\be
\int \de t\; \mathcal{P}_r=\int \de t\;\dot{h}_{ij} \partial_r h_{ij}=-i\int \de \omega\; \omega \varepsilon_{ij}^{*}\partial_r\varepsilon_{ij}.
\ee
Then, using Eq.~(\ref{eq:SolutionEpsilon}), we find
\be
\int \de t\; \mathcal{P}_r=\int \de \omega\; \frac{\kappa^2\omega^2}{16\pi^2 r^2}|t_{ij}|^2+ \mathcal{O}\left(\frac{1}{r^3}\right).
\ee
Plugging this result back into Eq.~(\ref{eq:EnergyRadiated}) and neglecting higher order terms we get
\be\label{eq:EnergyRadiated2}
E_{\textrm{rad}}=\frac{\kappa^2}{16\pi^2}\int \de \omega\; \omega^2 \oiint\de\Omega\;|t_{ij}|^2.
\ee
Let us now evaluate the modulus square which appears in the integrand. From Eq. \eqref{tij3}, we get the modulus square
\be
|t_{ij}|^2=\frac{\omega^4}{4}P_{il}M^{*}_{lm}P_{mj}P_{ik}M_{kq}P_{qj}=\frac{\omega^4}{4}M^{*}_{lm}P_{mj}M_{jk}P_{kl}.
\ee
In the last step we used the properties of the projection operator $P_{ij}=P_{ji}$ and $P_{il}P_{lj}=P_{ij}$ and the symmetry $M_{ij}=M_{ji}$. The evaluation is now straightforward, though a bit tedious, and the result is
\be
|t_{ij}|^2=\frac{\omega^4}{4}\left(| I_{ij}|^2-2~\hat{r}\cdot I^{*}\cdot I \cdot \hat{r}+\frac{1}{2}|\hat{r}\cdot I \cdot\hat{r}|^2\right),
\ee
where by dot we mean the standard matrix multiplication as in
\be
\hat{r}\cdot I \cdot\hat{r}=\hat{r}_pI_{pq}\hat{r}_q.
\ee
Thus, going back to Eq.~(\ref{eq:EnergyRadiated2}) we have
\be\label{eq:EnergyRadiated3}
E_{\textrm{rad}}=\frac{\kappa^2}{16\pi^2}\int \de \omega\; \frac{\omega^6}{4} \oiint\de\Omega\;\left[|I_{ij}|^2-2~\hat{r}\cdot I^{*}\cdot I \cdot \hat{r}+\frac{1}{2}|\hat{r}\cdot I \cdot\hat{r}|^2\right].
\ee
We now introduce the average over the entire solid angle, which we will denoted by $\langle ~\rangle\equiv\frac{1}{4\pi}\oiint\de\Omega\;$, and assume isotropy of the sources. The following properties will be useful
\begin{align}
\langle\hat{r}_i\hat{r}_j\rangle&=\frac{1}{3}\delta_{ij},\\
\langle\hat{r}_i\hat{r}_j\hat{r}_k\hat{r}_l\rangle&=\frac{1}{15}\left(\delta_{ij}\delta_{kl}+\delta_{ik}\delta_{jl}+\delta_{il}\delta_{jk}\right).
\end{align}
They imply
\begin{align}
\langle\hat{r}\cdot I^{*}\cdot I \cdot \hat{r}\rangle&=\frac{|I_{ij}|^2}{3},\\
\langle|\hat{r}\cdot I \cdot\hat{r}|^2\rangle&=\frac{2}{15}|I_{ij}|^2.
\end{align}
Hence Eq.~(\ref{eq:EnergyRadiated3}) reduces to the simple expression
\be\label{eq:EnergyRadiatedSimple}
E_{\textrm{rad}}=\frac{G}{5}\int \de \omega\; \omega^6 |I_{ij}|^2.
\ee
However, the last equation makes reference to quantities computed in the frequency domain, whereas we would like to go back to the time domain. For this purpose we introduce the anti-transform of the tensor $I_{ij}$
\be
\widetilde{I}_{ij}=\int\frac{\de\omega}{\sqrt{2\pi}}\; \e^{i\omega t}I_{ij},
\ee
whose second derivative with respect to time reads
\be
\frac{\de^2}{\de t^2}\widetilde{I}_{ij}=-\int \frac{\de \omega}{\sqrt{2\pi}}\;\e^{i\omega t}\omega^2I_{ij}=-2\int \frac{\de \omega}{\sqrt{2\pi}}\int\de^3\mbox{r}^{\prime}\;\e^{i\omega( t- \hat{r}\cdot r^{\prime})}\left(\tau_{ij}-\frac{1}{3}\delta_{ij}\tau_{kk}\right).
\ee
In the long wavelength limit $|\omega r^{\prime}|\ll1$, valid when the wavelength is much smaller than the spatial extension of the sources,
we can neglect the spatial dependence in the phase of the integrand and write, after inverting the Fourier transform of $\tau_{ij}$,
\be\label{eq:ItildeDoubleDot}
\frac{\de^2}{\de t^2}\widetilde{I}_{ij}=-2\int\de^3\mbox{r}^{\prime}\left(T_{ij}-\frac{1}{3}\delta_{ij}T_{kk}\right).
\ee
The last one is a useful relation, since the integral over time has disappeared and we are left with an integral of the stress-energy tensor over the spatial domain occupied by the sources. Yet it can be given a more convenient form, where only derivatives of the energy density appear, by using identities which follow from the conservation of the stress-energy tensor. In fact, we have
\be
\partial_l\partial_k T_{kl}=\ddot{T}_{00},
\ee
and therefore
\be\label{eq:integralidentitySources}
\int\de^3\mbox{r}^{\prime}\; r^{\prime}_i r^{\prime}_j \ddot{T}_{00}=\int\de^3\mbox{r}^{\prime}\; r^{\prime}_i r^{\prime}_j \partial_k\partial_l T_{kl}=2\int\mbox{d}^3 r^{\prime}\; T_{ij}.
\ee
In the last step, integration by parts and reshuffling of indices have been performed. Substituting identity Eq.~(\ref{eq:integralidentitySources}) into Eq.~(\ref{eq:ItildeDoubleDot}) we get
\be\label{eq:SecondDerQuadrupole}
\frac{\de^2}{\de t^2}\widetilde{I}_{ij}=-\int\de^3\mbox{r}^{\prime}\left(r^{\prime}_ir^{\prime}_j-\frac{1}{3}(r^{\prime})^2\delta_{ij}\right)\ddot{T}_{00}.
\ee
The last formula is an important one since we recognise the mass quadrupole moment in the rhs. In fact as a consequence of 
energy-momentum conservation, the leading contribution to the emission of gravitational waves arises from the quadrupole moment in a multipole expansion of the sources. More precisely, the monopole term vanishes due to mass-energy conservation, whereas the vanishing of the dipole term follows from the conservation of linear and angular momentum \cite{Misner:1974qy}. This leads to a restriction of the class of physical systems where gravitational waves emission can be observed, at the same time strongly constraining their intensity. The last statement will be made clearer from the quadrupole formula given below. Computing the $L^2\mbox{-norm}$ of the third derivative of $\widetilde{I}_{ij}$ and using the Parseval's theorem, we obtain
\be
\int\de t\; \left|\frac{\de^3}{\de t^3}\widetilde{I}_{ij}\right|^2=\int\de\omega\;\omega^6 \left|I_{ij}\right|^2.
\ee
Finally, we can write the gravitational radiation emitted by isotropic sources in the Newtonian limit as
\be\label{eq:EnergyLossFinal}
E_{\textrm{rad}}=\frac{G}{5}\int\de t\; \dddot{\widetilde{I}}_{ij}\dddot{\widetilde{I}}_{ij}.
\ee
A more complete analysis (see \emph{e.g.} \cite{Misner:1974qy}) reveals that the power emitted instantaneously is given by the integrand of Eq.~(\ref{eq:EnergyLossFinal})\footnote{The cautious reader might have noticed that in Eq.~(\ref{eq:EnergyLossFinal}) the integral is evaluated over the whole time axis. Therefore a priori we cannot naively identify its integrand with the power.}
\be\label{eq:quadrupole}
\frac{\de E_{\textrm{rad}}}{\de t}=\frac{G}{5} \dddot{\widetilde{I}}_{ij}\dddot{\widetilde{I}}_{ij}.
\ee
This is the well-known quadrupole formula.

\newpage

\section{Newtonian binaries} \label{Sec 6}
An interesting and immediate application of the theory of emission of gravitational waves developed above is the Newtonian binaries, which are non-relativistic systems of two stars orbiting around their center of mass, as shown in Fig.~2.
\begin{figure}[h!]
\centering
\def\svgscale{0.8}
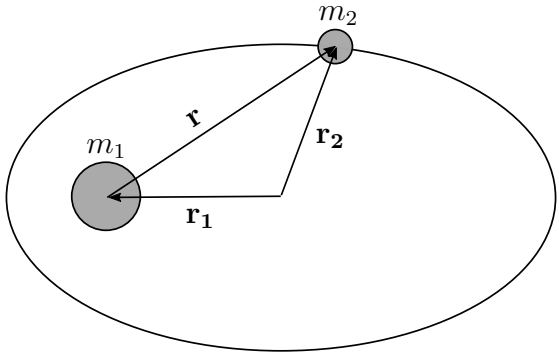
\caption{Binary system governed by Newtonian gravity.}
\end{figure}

Given two massive point-like bodies in the Newtonian regime, their dynamics is completely described by the reduced one-body problem:
\be
\ddot{\mathbf{r}}=-\frac{G M}{r^3}\mathbf{r},
\ee
where $\mathbf{r}=\mathbf{r}_2-\mathbf{r}_1$ is the position of the second body relative to the first one and $M=m_1+m_2$ is the total mass of the binary system. This equation admits circular orbits as its solutions
\be
\mathbf{r}=r\left( \cos\omega t,\sin\omega t,0\right).
\ee
The orbits lie in the equatorial plane, by conservation of angular momentum.
Choosing a frame where the origin of the axes coincides with the centre of mass, one has
\begin{align}
\mathbf{r}_1=-\frac{\mu r}{m_1}\left( \cos\omega t,\sin\omega t,0\right),\\
\mathbf{r}_2=\frac{\mu r}{m_2}\left( \cos\omega t,\sin\omega t,0\right),
\end{align}
where we introduced the reduced mass
\be
\mu=\frac{m_1 m_2}{M}.
\ee
The angular velocity $\omega$, according to Kepler law, is given by
\be \label{angular}
\omega^2=\frac{G M}{r^3}.
\ee
The quadrupole moment of the mass distribution is given by
\be
Q_{ij}=\sum_{a=1,2}m_a\left(r_{ai}r_{aj}-\frac{1}{3}\delta_{ij}r_a^2\right).
\ee
Since the positions of the two bodies evolves with time, $Q_{ij}$ is also a function of time. Its derivatives of order two or higher coincide with those of $-\widetilde{I}_{ij}$, as shown by Eq.~(\ref{eq:SecondDerQuadrupole}). One finds, writing rank-two tensors in matrix form, that 
\be
\left(Q_{ij}\right)=\mu^2 r^2 \left(\frac{1}{m_1}+\frac{1}{m_2}\right)\begin{pmatrix} \cos^2\omega t -\frac{1}{3} & \cos\omega t\sin\omega t & 0\\
 \sin\omega t\cos\omega t & \sin^2\omega t -\frac{1}{3} & 0\\ 0& 0& -\frac{1}{3}\end{pmatrix},
\ee
whence it follows
\be
\left(\dddot{\widetilde{I}_{ij}}\right)=4\mu r^2 \omega^3 \begin{pmatrix} \sin2\omega t & -\cos2\omega t & 0\\
-\cos2\omega t & -\sin2\omega t  & 0\\ 0& 0& 0\end{pmatrix}.
\ee
Therefore, using Eq.~(\ref{eq:quadrupole}), we can write for the energy lost by the system in the form of gravitational waves in a time unit
\be
\frac{\de E_{\textrm{rad}}}{\de t}= \frac{32\mu^2M^3 G^4}{5 r^5}.
\ee
To get the order of magnitude involved for Newton binary system, let us use Eq. \eqref{angular} to write the frequency as
\begin{equation}
f \equiv  \frac{\omega}{\pi} = (1.16 \times 10^{-7} \mbox{Hz}) \sqrt{\frac{M}{M_{sun}} \left(\frac{1 \hspace{1mm} \mbox{AU}}{r}\right)^3},
\end{equation}
where we adopted $M_{sun} = 3.0 \times 10^{30} \hspace{1mm}\mbox{Kg}$ and $1 \hspace{1mm} \mbox{AU} = 1.5 \times 10^{11}\hspace{1mm} \mbox{m}$.
Assuming that $m_1=m_2=M_{sun}/2$ and $r = 1 \hspace{1mm} \mbox{AU} $, we have a frequency of $ 0.58 \times 10^{-7}\;\mbox{Hz}$ and an emitted power of $1.4 \times 10^{12} \hspace{1mm} \mbox{W}$.

\begin{figure}[t]
\begin{center}
\includegraphics[scale=0.5]{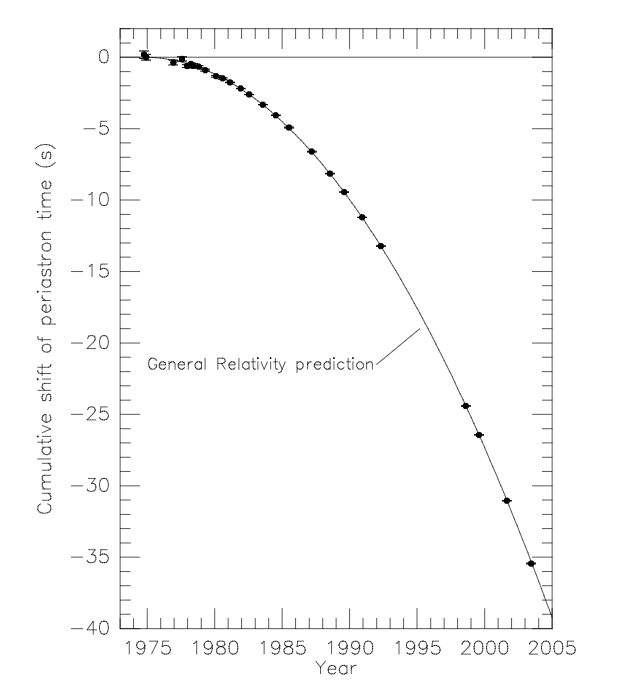}
\caption{Cumulative periastron shift of the binary pulsar PSR 1913+16 {\em\cite{Weisberg:2004hi}}.}
\label{cumulative shift}
\end{center}
\end{figure}

In 1975, Hulse and Taylor\footnote{They were awarded the 1993 Nobel Prize for Physics for their discovery.} \cite{Hulse:1974eb} discovered a binary system constituted by a pulsar, named PRS 1913+16, with mass $m_1=1.44 \; M_{sun}$ and a companion neutron star of mass $m_2 = 1.39 \; M_{sun}$. The orbital period of the binary system was 7 hours 45 min 7s, but over many years of observations, astronomers observed that such orbital period was slightly decreasing and the two neutron stars were spiralling into one another. Weisberg and Taylor \cite{Weisberg:2004hi} recognised that this phenomenon was due to the emission of energy in the form of gravitational waves, and they showed that the decrease in the orbital period was in agreement with the predictions of General Relativity. Fig.~\ref{cumulative shift} shows the shift in the periastron due to to the changing size of the orbit radius. 

\newpage~\newpage
\section{Outlook}

In these lectures notes we have presented a field theoretical approach to the construction of General Relativity, starting 
from free tensor fields of spin $s = 2$, and leading after summing infinitely many interaction terms to the 
action (\ref{2.16}) originally presented by Einstein \ct{Einstein:1916vd}. In this approach the geometric interpretation
of gravity arises from the field theory, instead of being the starting point of the construction of the theory. We went on 
to discuss the interpretation of free massless tensor fields as an approximation of gravitational waves and considered 
their emission by sources with time-varying quadrupole moment, applying it to simple Newtonian binaries. It is 
straightforward, though not simple, to carry this approach further by considering waves on curved background 
space-times like black holes \ct{Wheeler:1957,Zerilli:1970}. An application is the emission of gravitational waves by compact 
objects orbiting a black-hole as discussed for example in \ct{Martel:2005,Koekoek:2011, dAmbrosi:2015}. 

Relativistic corrections to binary stars like the Hulse-Taylor system are customarily obtained by computing 
post-newtonian corrections to Kepler orbits and Minkowski space-time. A detailed review can be found in 
\ct{Maggiore:2008}. The most exciting development in recent times was the actual discovery of direct gravitational 
wave signals by the LIGO and Virgo consortium \ct{Abbott:2016blz}, shortly after these lectures were given, 
resulting from the merger of two rather massive black holes. Modelling such systems requires in addition to 
a more sophisticated theoretical approach also extensive numerical computations. Many more sources of 
gravitational waves are expected in the near future with the advanced LIGO and Virgo interferometers, and 
in the more distant future with new instruments like the proposed Einstein Telescope and the spaceborn eLISA 
observatory.

\section*{Acknowledgments}
These notes are based on the lectures given by one of us (van Holten) at the 21st edition of the W.E. Heraeus Summer School ``Saalburg" for Graduate Students on ``Foundations and New Methods in Theoretical Physics" in September 2015.
We are grateful to the organisers of the school for providing the opportunity to discuss and elaborate these topics in a stimulating atmosphere. 
RO acknowledges the current support of the ERC Starting Grant 335146 ``HoloBHC''.
The work of JWvH is carried out as part of the research programme of NWO-I on Gravitational Physics. MdC would
like to thank Thomas Helfer and Mairi Sakellariadou for useful comments on the manuscript. RO is grateful to Geoffrey Comp{\`e}re, Fabrizio Finozzi and Federico Mogavero for valuable feedback. Finally, the authors thank Stanley Deser for comments on the manuscript.

\newpage

\bibliography{references}

\end{document}

%% file: figures/gw.pdf_tex
\begingroup%
  \makeatletter%
  \providecommand\color[2][]{%
    \errmessage{(Inkscape) Color is used for the text in Inkscape, but the package 'color.sty' is not loaded}%
    \renewcommand\color[2][]{}%
  }%
  \providecommand\transparent[1]{%
    \errmessage{(Inkscape) Transparency is used (non-zero) for the text in Inkscape, but the package 'transparent.sty' is not loaded}%
    \renewcommand\transparent[1]{}%
  }%
  \providecommand\rotatebox[2]{#2}%
  \ifx\svgwidth\undefined%
    \setlength{\unitlength}{401.25323777bp}%
    \ifx\svgscale\undefined%
      \relax%
    \else%
      \setlength{\unitlength}{\unitlength * \real{\svgscale}}%
    \fi%
  \else%
    \setlength{\unitlength}{\svgwidth}%
  \fi%
  \global\let\svgwidth\undefined%
  \global\let\svgscale\undefined%
  \makeatother%
  \begin{picture}(1,0.32571714)%
    \put(0,0){\includegraphics[width=\unitlength,page=1]{gw.pdf}}%
    \put(0.4459572,0.20938299){\color[rgb]{0,0,0}\makebox(0,0)[lb]{\smash{\textit{region $\Sigma$}}}}%
    \put(-0.000237,0.15433681){\color[rgb]{0,0,0}\makebox(0,0)[lb]{\smash{\textit{surface of integration $S$}}}}%
    \put(0,0){\includegraphics[width=\unitlength,page=2]{gw.pdf}}%
    \put(0.77076322,0.30459699){\color[rgb]{0,0,0}\makebox(0,0)[lt]{\begin{minipage}{0.35887561\unitlength}\raggedright \end{minipage}}}%
    \put(0.7146019,0.27516953){\color[rgb]{0,0,0}\makebox(0,0)[lb]{\smash{\textit{flux of gravitational energy}}}}%
    \put(0,0){\includegraphics[width=\unitlength,page=3]{gw.pdf}}%
  \end{picture}%
\endgroup%

%% file: figures/binsystem.pdf_tex
\begingroup%
  \makeatletter%
  \providecommand\color[2][]{%
    \errmessage{(Inkscape) Color is used for the text in Inkscape, but the package 'color.sty' is not loaded}%
    \renewcommand\color[2][]{}%
  }%
  \providecommand\transparent[1]{%
    \errmessage{(Inkscape) Transparency is used (non-zero) for the text in Inkscape, but the package 'transparent.sty' is not loaded}%
    \renewcommand\transparent[1]{}%
  }%
  \providecommand\rotatebox[2]{#2}%
  \ifx\svgwidth\undefined%
    \setlength{\unitlength}{256.8bp}%
    \ifx\svgscale\undefined%
      \relax%
    \else%
      \setlength{\unitlength}{\unitlength * \real{\svgscale}}%
    \fi%
  \else%
    \setlength{\unitlength}{\svgwidth}%
  \fi%
  \global\let\svgwidth\undefined%
  \global\let\svgscale\undefined%
  \makeatother%
  \begin{picture}(1,0.63642798)%
    \put(0,0){\includegraphics[width=\unitlength,page=1]{binsystem.pdf}}%
    \put(0.14659753,0.36334143){\color[rgb]{0,0,0}\makebox(0,0)[lb]{\smash{$m_1$}}}%
    \put(0.56796769,0.60684209){\color[rgb]{0,0,0}\makebox(0,0)[lb]{\smash{$m_2$}}}%
    \put(0.32727117,0.23507505){\color[rgb]{0,0,0}\makebox(0,0)[lb]{\smash{$\bold{r_1}$}}}%
    \put(0.56317386,0.38147225){\color[rgb]{0,0,0}\makebox(0,0)[lb]{\smash{$\bold{r_2}$}}}%
    \put(0,0){\includegraphics[width=\unitlength,page=2]{binsystem.pdf}}%
    \put(0.34000164,0.40924093){\color[rgb]{0,0,0}\rotatebox{33.48335374}{\makebox(0,0)[lb]{\smash{$\bold{r}$}}}}%
  \end{picture}%
\endgroup%